%% file: main_arxiv.tex
\documentclass[letterpaper,twocolumn,10pt]{article}
\usepackage{usenix-2020-09}

\usepackage{tikz}
\usepackage{filecontents}

\usepackage{algorithm}
\usepackage{algorithmic}
\usepackage{array}
\usepackage{multirow}
\usepackage{caption}
\usepackage{adjustbox}
\usepackage{cellspace}
\usepackage{tabularx}
\usepackage{hhline}
\usepackage{makecell}
\usepackage{subcaption}
\usepackage{mdwmath}
\usepackage{mdwtab}
\usepackage{eqparbox}
\usepackage{stfloats}
\usepackage[cmex10]{amsmath}
\usepackage{bm} 
\usepackage{amssymb}
\usepackage{todonotes}

\begin{document}

%don't want date printed
\date{}

\title{\Large \bf MARS: Defending Unmanned Aerial Vehicles From Attacks on Inertial Sensors with Model-based Anomaly Detection and Recovery}

\author{
{\rm Haocheng Meng}\\
Duke University
% haocheng.meng@duke.edu
\and
{\rm Shaocheng Luo}\\
Duke University
% shaocheng.luo@duke.edu
\and
{\rm Zhenyuan Liang}\\
Duke University
% zhenyuan.liang@duke.edu
\and
{\rm Qing Huang}\\
Duke University
% jim.huang@duke.edu
\and
{\rm Amir Khazraei}\\
Duke University
% amir.khazraei100@gmail.com
\and
{\rm Miroslav Pajic}\\
Duke University
% miroslav.pajic@duke.edu
% copy the following lines to add more authors
% \and
% {\rm Name}\\
%Name Institution
} % end author

\maketitle

\begin{abstract}
Unmanned Aerial Vehicles (UAVs) rely on measurements from Inertial Measurement Units (IMUs) to maintain stable flight. However, IMUs are susceptible to physical attacks, including acoustic resonant and electromagnetic interference attacks, resulting in immediate UAV crashes. Consequently, we introduce a \emph{Model-based Anomaly detection and Recovery System} (MARS) that enables UAVs to quickly detect adversarial attacks on inertial sensors and achieve dynamic flight recovery. MARS features an attack-resilient state estimator based on the Extended Kalman Filter, which incorporates position, velocity, heading, and rotor speed measurements to reconstruct accurate attitude and angular velocity information for UAV control. Moreover, a statistical anomaly detection system monitors IMU sensor data, raising a system-level alert if an attack is detected. Upon receiving the alert, a multi-stage dynamic flight recovery strategy suspends the ongoing mission, stabilizes the drone in a hovering condition, and then resumes tasks under the resilient control. Experimental results in PX4 software-in-the-loop environments as well as real-world MARS-PX4 autopilot-equipped drones demonstrate the superiority of our approach over existing IMU-defense frameworks, showcasing the ability of the UAVs to survive attacks and complete the missions.
\end{abstract}

\input{Introduction}
\input{Preliminaries}

\input{Methodology}

\input{Evaluation}

\input{Experiment}
\input{Discussion}

% \section*{Acknowledgments}
% %-------------------------------------------------------------------------------

% %-------------------------------------------------------------------------------
\twocolumn[\newpage]
\section*{Ethics Considerations}

We have identified multiple stakeholders for this research: general UAV users for commercial or industrial purposes; researchers in UAV security; potential malicious parties who could misuse the identified sensor attacks. For the vulnerabilities discussed in this research, they are well-known attacks identified in prior works. Our work does not introduce new threats but focuses on mitigating the known ones. Additionally, we ensure no damage are made from our simulations and experiments, as they were conducted in controlled and isolated environments without interacting with live systems or real-world UAV users. After publication, we will reach out to work with UAV manufacturers incorporating our defenses into the latest UAV developments. Therefore, we consider it ethical and beneficial to introduce our work to the community to improve overall UAV resilience and prevent potential damage to UAV users and the general public.

% %-------------------------------------------------------------------------------
\section*{Open Science}
We adhere to the open science policy by making the source code, datasets and demonstrations for both simulations and real-world experiments available on our project website\cite{MARS}.

\section*{Acknowledgment}
This work is sponsored in part by the ONR under agreement N00014-23-1-2206, AFOSR under the award number FA9550-19-1-0169, and by the NSF under NAIAD Award 2332744 as well as the National AI Institute for Edge Computing Leveraging Next Generation Wireless Networks, Grant CNS-2112562.

% USENIX program committees give extra points to submissions that are
% backed by artifacts that are publicly available. If you made your code
% or data available, it's worth mentioning this fact in a dedicated
% section.

%-------------------------------------------------------------------------------
% \bibliographystyle{plain}
% \twocolumn[\newpage]
\bibliographystyle{unsrt}
\bibliography{reference,CPSL_DukePapers}
%%%%%%%%%%%%%%%%%%%%%%%%%%%%%%%%%%%%%%%%%%%%%%%%%%%%%%%%%%%%%%%%%%%%%%%%%%%%%%%%
\input{Appendix}

\end{document}

%% file: Introduction.tex
\section{Introduction}
\label{sec:intro}

Inertial Measurement Units (IMUs) are used in any Unmanned Aerial Vehicle (UAV) to maintain the flight stability. IMU incorporates the orientation, angular velocity, and acceleration from the accelerometer, gyroscope, and magnetometer, providing vital information on body-frame acceleration and attitude to keep the drone flying smoothly (e.g.,~\cite{martin2010true}). However, UAV inertial sensors have been shown vulnerable to various adversarial attacks. Onboard gyroscopes and accelerometers are typically Micro-Electro-Mechanical Systems (MEMS) devices, %chosen for their size, cost-efficiency, and adequate performance, 
which makes them sensitive to acoustic noise at resonant frequencies~\cite{castro2007influence}. While the resonant frequencies are typically in the ultrasonic band, commercially available IMU sensors can resonate within the audible band, making attacks feasible using consumer-grade speakers. Specifically, the resonant oscillation caused by acoustic injection can severely disrupt the drone's attitude controllers~\cite{son2015rocking, kim2024systematic}. Moreover, attackers can gain implicit control over the output of compromised sensors by manipulating the injected analog signals~\cite{tu2018injected}. 

IMUs are also vulnerable to Electromagnetic Interference (EMI) attacks. EMI waves typically require significant power %consumption 
to distort the circuits~\cite{kune2013ghost,backstrom2004susceptibility}. %However, it has been shown that IMU sensor circuits are susceptible to intentional electromagnetic interference~\cite{delsing2006susceptibility}. 
% Recently, it has been shown that 
Recent EMI attacks on commercial flight controller boards have been shown to incapacitate UAVs by corrupting the communication between IMU sensors and the flight control unit~\cite{jang2023paralyzing}.

Both types of attacks result in safety-critical incidents, as a UAV is rendered uncontrollable under the inertial sensor attacks, crashing within seconds; thus, igniting recent interest in defenses against IMU attacks~\cite{tu2019flight,choi2020software, fei2020learn, zhang2020real, akowuah2021recovery, jeong2023rocking}.
Yet, existing prevention and mitigation techniques fail to neutralize the impact of attacks; e.g., \cite{choi2020software, fei2020learn, zhang2020real, akowuah2021recovery} fail against IMU attacks~\cite{son2015rocking,tu2018injected,jang2023paralyzing} that quickly crash the UAVs (see~\cite{jang2023paralyzing} for details). Installing redundant IMUs is also ineffective, as attackers would compromise all IMUs that share the same physical characteristics and communication protocols. Shielding methods are constrained by the physical properties of drone sensors and control boards; metal shields can interfere with wireless communication, increase the risk of overheating, degrade performance, and reduce payload capacity due to their weight (e.g.,~\cite{doriol2009emc, geetha2009emi}). 

Thus, filtering or de-noising compromised IMU readings has been considered (e.g.,~\cite{karaim2019low, 8836664}). %~\cite{nirmal2016noise, karaim2019low, 8836664}. 
Traditional frequency-based filters struggle because attack signals can disperse across the in-band domain, so Deep Auto Encoders (DAE) have been deployed to learn the acoustic resonance signature~\cite{jeong2023rocking}; however, as we show in this work, 
they are unable to cope with sensor saturation, unexpected sensor measurements, and communication drops. Moreover, they lack real-time guarantees for online recovery~\cite{jeong2023rocking}.

Instead of attempting to de-noise the compromised IMU measurements, a complementary attitude reconstruction method leverages geometric relationships to restore attitude control~\cite{tu2019flight}. Yet, this approach requires unrealistically high accuracy and a high update frequency (as fast as the IMU sampling rate) of the position sensors; this is unfeasible in real-world scenarios where GPS is usually sampled 10 times slower than IMUs (e.g., $5\sim 10~Hz$ vs. $100~Hz$). %\todo{Luo: our experiments used GPS/Vicon at 10-20Hz, and a typical IMU frequency is more than 100Hz, shall we change it to be $10~Hz$ vs. $100~Hz$?} 

% In view of the severity of threats to UAV onboard IMUs and the limitations of existing recovery methods

Consequently, in this work, we introduce a \emph{Model-based Anomaly detection and Recovery System} (MARS) to provide resiliency for UAVs against physical attacks on the inertial sensors. By utilizing tachometers that measure propeller rotational speeds, we identify an alternative solution to UAV body-frame state estimation, independent of IMUs. Tachometers are small, very low-cost sensors consisting of pulse counters and sensing probes, which can be easily installed on UAV arms. These sensors have optical sensing and magnetic sensing mode, and both cannot be compromised with acoustic and electromagnetic interference. As encoders, they can measure rotor speeds with high resolution and frequency, sufficient for real-time estimation~and~control~\cite{6631474}.

By exploiting our knowledge (i.e., the model) of the physical dynamics of a UAV, we~find~that \emph{rotor speed measurements can be combined with existing non-IMU position and heading data to estimate the net thrust and torque applied at the UAV's center of gravity}. This information is then used to produce body-frame attitude estimates through a \textit{novel non-linear sensor fusion}. This approach offers resiliency to IMU attacks from two perspectives: first, it can be integrated with anomaly detection methods to accurately and swiftly detect IMU attacks and trigger system-level alarms; second, it enables position and attitude control in the absence of trustworthy IMU data. Based on these findings, we design a \emph{flight recovery strategy} that effectively handles adversarial situations in UAV missions by detecting attacks, restoring stability, and continuing missions with %enhanced security and 
minimal performance degradation. 

In brief, MARS combines resilient sensor fusion, anomaly detection, and flight recovery, thereby safeguarding UAVs from corrupted inertial sensor measurements and maintaining an adequate level of the UAV flight performance. 
Our contributions can be summarized as:

% \begin{itemize}%[leftmargin=*]
% % \begin{itemize}[leftmargin=12pt]\setlength\itemsep{-4pt}
%     % \item We introduce a general attack-resilient state estimator that does not require IMUs to estimate the applied UAV control inputs (i.e., thrust and torque), and incorporates position and heading sensors to generate high-fidelity body-frame orientation and angular velocity estimates.
\vspace{2pt}
\textbullet~We introduce a general attack-resilient state estimator that does not require IMUs but incorporates position, heading and tachometer sensors to estimate {net thrust and torque}, and then generate high-fidelity body-frame orientation and angular velocity~estimates.

\vspace{2pt}
\textbullet~We design a MARS-based CUSUM sliding window anomaly detector (AD) that significantly outperforms existing ADs (e.g.,~\cite{ye2001anomaly, quinonez2020savior}) by swiftly and accurately detecting IMU attacks and providing system-level alarms for UAV security-aware decision-making.

\vspace{2pt}
\textbullet~We develop a multi-stage flight recovery strategy that prioritizes security without significantly sacrificing performance. Our method is universally applicable to arbitrary UAV missions, regardless of the UAV's state, ensuring freedom in UAV maneuverability when equipped with the MARS detection and recovery framework.

\vspace{2pt}
\textbullet~We evaluate MARS methodology on the PX4 open-source autopilot firmware~\cite{7140074, PX4} in large-scale simulations as well as real-world physical experiments; we demonstrate MARS effectiveness, even against attacks for which other security methods (e.g.,~\cite{tu2019flight,jeong2023rocking}) fail. 
% \end{itemize}

The paper is organized as follows. Sec.~\ref{sec:prelim} introduces preliminaries on nonlinear UAV models, IMU vulnerabilities and other %relevant 
related work. Sec.~\ref{sec:mars} presents the MARS methodology, based on a novel model-based resilient state estimator, MARS-based anomaly detection, and the multi-stage flight recovery. % strategy. 
Sec.~\ref{sec:eval1} presents large-scale~evaluation in a simulated environment, compares %the proposed method 
MARS with existing frameworks in a hovering mission and explores the efficacy of flight recovery in a dynamic tracking mission. Sec.~\ref{sec:physical} presents real drone case studies with our customized MARS-PX4 autopilot and %further demonstrates 
demonstrating MARS %efficacy of our proposed method 
effectiveness in real-world implementations. %Finally, we 
% We discuss limitations and future work %avenues %for future work 
% (Sec.~\ref{sec:future}), before concluding remarks~(Sec.~\ref{sec:conclusion}).
We provide concluding remarks and discuss MARS limitations and future work~%for future work 
(Sec.~\ref{sec:future}). 

%% file: Preliminaries.tex
\section{Preliminaries}
\label{sec:prelim}

\input{Related_Work}

%% file: Related_Work.tex
\subsection{UAV Modeling and Control}

\noindent\textbf{Nonlinear UAV/quadrotor model.} %The quadcopter system can be considered 
UAVs are modeled as a nonlinear 6-Degree of Freedom (6-DoF) rigid body system with 3 translational and 3 rotational movements, utilizing 4 rotors to generate aerodynamic force for flight~\cite{khan2014quadcopter}. We consider two reference frames: % in our discussion: 
the \emph{Earth} (inertial frame) and the \emph{body} frame. The Earth frame is defined as the right-handed North, East, Down (NED) frame, denoted by $\mathcal{F}_{\mathcal{E}}$. The right-handed body frame, defined as Forward, Right, Down (FRD) frame, is denoted by $\mathcal{F}_{{\mathcal{B}}}$. The rotation from the Earth %frame 
to the body frame is captured by the rotation matrix $\mathbf{R}(\mathbf{q})\in\mathbb{R}^{ 3\times 3}$, characterized by quaternion $\mathbf{q} =\begin{bmatrix}q_w, q_x, q_y, q_z\end{bmatrix}^T$. The translational motion is captured by the Earth frame positions expressed as $\mathcal{\xi}=\begin{bmatrix}x,y,z\end{bmatrix}^T$. The rotor speeds of the four propellers are given as $\mathbf{\omega}=\begin{bmatrix}\omega_1, \omega_2, \omega_3, \omega_4\end{bmatrix}^T$. App.~\ref{appendix:drone_model} provides detailed description of the nonlinear model capturing the UAV physics (i.e., its \emph{physical} dynamics).

\vspace{3.2pt}
\noindent\textbf{Onboard sensing.} A UAV relies on onboard sensors to %perform sensor fusion and 
achieve stable and precise flight control. % in autonomous missions. 
The key %sensing components %of a quadcopter system 
sensors %include 
are IMU, compass, barometer, and GPS, which provide real-time attitude and position information. IMU units typically consist of accelerometers, gyroscopes, and magnetometers, measuring changes in the drone's linear velocity, angular velocity, and magnetic field. Also, sensors such as optical flow, depth cameras, radar, and Lidar can further improve the UAV's situational-awareness. In this paper, we focus on the two major components inside the IMU--accelerometers and gyroscopes, as well as the GPS and compass. These sensors produce body-frame attitude and angular velocity, earth-frame position, and heading measurements.

\vspace{3.2pt}
\noindent\textbf{Standard UAV state-estimators.}
%The design of a s
A standard UAV state estimator, used for fusion of the multi-modal sensing, is built upon GPS-IMU sensor fusion, where the IMU is \emph{currently irreplaceable} due to its direct observation of fast-changing translational and rotational~motions~\cite{savage1998strapdown1, savage1998strapdown2}.

\begin{figure}[!t]
\centerline{\includegraphics[width=0.908\columnwidth]{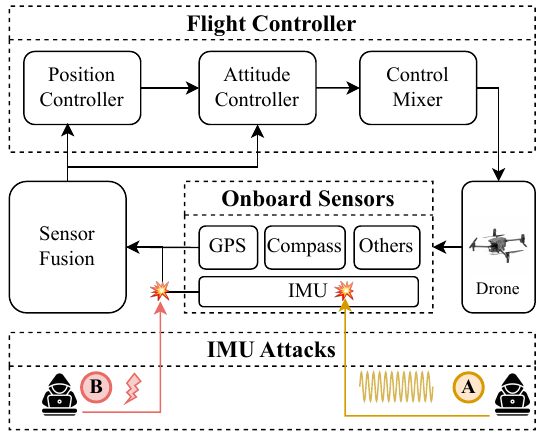}}
\caption{UAV onboard sensing and control architecture highlighting physical vulnerabilities of the %inertial 
IMUs:~A)~Acoustic resonant; B)~Electromagnetic interference~attack.} 
\label{fig:standard_control_overview}
\end{figure}

\vspace{3.2pt}
\noindent\textbf{Cascade position and attitude flight controllers.}
With accurate position and attitude estimates from an IMU-based state estimator, an efficient flight controller calculates the desired rotor speed commands for the four rotors. A commonly used control architecture is the cascaded position and attitude PID control: the position controller takes the position set-points provided by high-level modules and the position estimates from a low-level filter, producing attitude and thrust commands. The attitude controller then computes the error between the desired attitude and the current estimate, yielding the desired body-frame torque and thrust, which are translated into rotor speed commands by the control mixer. Fig.~\ref{fig:standard_control_overview} illustrates the overall high-level sensing and control~architecture. 

%\subsection{Physical Attacks on IMUs and Defenses}
\subsection{Threat Model on IMU Attacks}
\label{sec:related}

% \textbf{Physical Attacks on UAVs.}
% The increasing utilization of UAVs expose them to a wide range of cyber-physical threats targeting onboard sensors. The malicious parties aim to directly or indirectly manipulate sensor data or control data thereby disrupting or taking over the normal flight and mission  execution of UAVs. 
In general, physical attack vectors on UAVs involve different spoofing properties, operations, victim characteristics and goals. They could affect GPS, LiDAR, Camera, IMU, audio and ultrasonic components %that are equipped
available on UAVs (e.g., see~\cite{xu2023sok}). %For example, GPS spoofing manipulates the true localization of the UAV and attempts to evade detection by the target GPS receiver and the target system’s state estimator~\cite{kerns2014unmanned}. Attack vectors on camera sensors compromise the perception of the environment of drones, disrupting the flight control~\cite{davidson2016controlling} and the high-level decision-making process~\cite{zhou2022doublestar}.
In this work, we focus on the reported inertial sensor vulnerabilities. 
% 
% \subsection{Inertial Sensor Vulnerabilities}
% 
Recently, several physical attacks have %been identified showing that attackers can 
compromised the IMU physical sensing mechanisms or disrupted communication pathways between the sensors and control units. Specifically, for the threat model from Fig.~\ref{fig:standard_control_overview}, adversaries can employ two \textbf{\emph{physical}} attacks---\emph{acoustic resonant attacks} or \emph{electromagnetic interference attacks}, 
to achieve the \textbf{\emph{attack objective}}: \emph{destabilize the low-level attitude control, leading to UAV crashes.} The following discusses the \emph{attacker knowledge and capability} of these two types of attacks.

\vspace{2pt}
\noindent\textbf{Acoustic resonant attacks.}
As MEMS devices, accelerometers and gyroscopes can experience resonance in their proof masses. Acoustic injection can induce resonant oscillations within these sensors, leading to significant fluctuations in the measurements. Attackers can exploit this inherent vulnerability %of MEMS devices 
by learning the acoustic resonance signature and identifying the susceptible range for successful attacks. The attacker can then propagate resonant signals using signal generators and transmitters to effectively trigger resonance in the UAV's onboard IMUs. A detailed description %on the properties 
of acoustic resonant attacks is provided in App.~\ref{subsec:appendix_acoustic}.

\vspace{2pt}
\noindent\textbf{Electromagnetic interference attacks.}
% \subsubsection{EMI attack methodology} 
Attackers can also leverage electromagnetic interference to affect the data transmission between the IMU sensors and the flight controller board, resulting in  communication failures~\cite{jang2023paralyzing}; these in turn, cause devastating estimation and control errors. The attack requirements, effect, and the comparison with acoustic resonant attacks are summarized in App.~\ref{subsec:appendix_emi}.

\vspace{2pt}
% \noindent\textbf{Anomaly Detection (AD) in UAVs.}
\subsection{Anomaly Detection (AD) in UAVs}
% As a response to the vulnerabilities in UAVs sensing, decision making and control, modern 
% Intrusion detection systems that are specific to UAVs have been developing thanks to the advancement in machine learning. The proper utilization of flight data is the core of efficient detection and mitigation, where model-based methods and data-driven methods are widely adopted to UAV systems~\cite{yang2023survey}. 
% 
Model-based and data-driven AD methods have been explored for UAVs~\cite{yang2023survey}. Model-based methods focus on the physical invariants of the drone dynamics~\cite{quinonez2020savior}, and have the advantage of simplicity and cost-efficiency~\cite{guo2017multisensor}, but require expert knowledge of the system dynamics and accurate estimation of system parameters~\cite{wang2016bias}. Data-driven methods exploit the hidden characteristics in sensor/control data, without the need for complicated modeling of the physical platform. The real-time model-free data analysis may provide fast and accurate detection of sensor faults~\cite{sun2017novel}, but requires significant training data to fully utilize the generality of neural~networks; otherwise, as we show, may not perform well.

%% file: Methodology.tex
\section%{Model-based Anomaly Detection and Recovery System (MARS)}
{MARS Methodology}
\label{sec:mars}

% \subsection{Overview}

\begin{figure*}[!t]
\centering
\includegraphics[width=\textwidth]{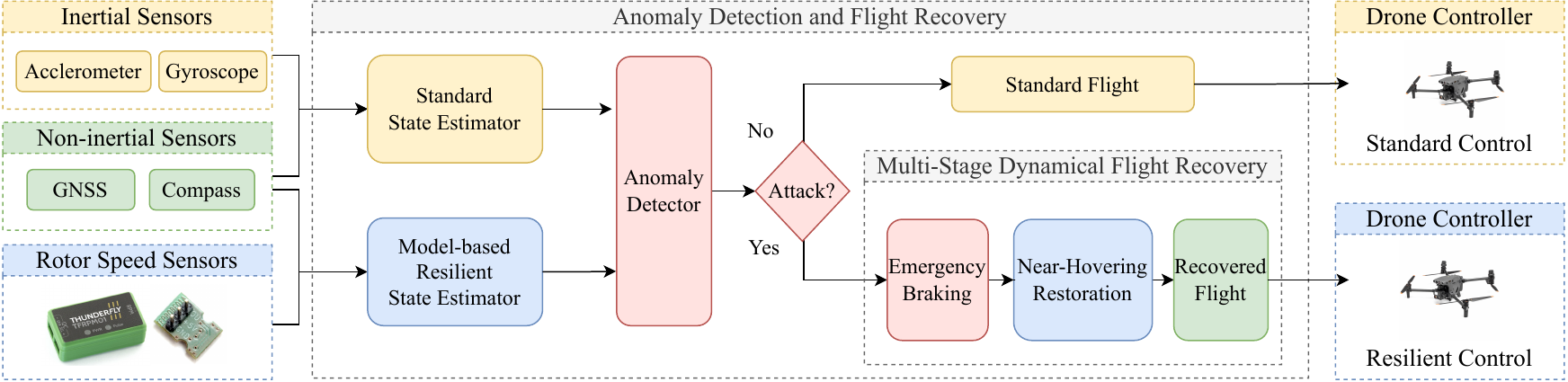}
\caption{MARS architecture with resilient sensor hubs (i.e., all sensors except the IMU), AD, and multi-stage~flight recovery, operating at two different modes: \emph{performance-focused standard control} and \emph{security-focused resilient control}~mode.}
\label{fig:mars}
\end{figure*}

MARS is a general anomaly detection and recovery framework that is universally applicable for defending UAVs %systems 
against cyber-physical threats targeting the onboard IMUs. %The idea behind MARS is to 
MARS 
uses the dynamical model (i.e., physics) of the UAV to 
quickly detect an attack on inertial sensors, and isolate the corrupted state estimation caused by the attacks as well as reconstruct the missing attitude information required by the UAV controllers; this is achieved by utilizing additional sensors that are unaffected by the IMU attack vectors. %\todo{Check: I haven't seen very relevant reference on this, shall we say "we discover that low-cost... can provide...?"} 
Specifically, %to find an appropriate replacement 
when inertial sensors are under attack, %we adopt 
additional low-cost rotor speed sensors (i.e., tachometers) may provide indirect body-frame attitude information using the unique UAV physical~dynamics. Fig.~\ref{fig:mars} shows the high-level architecture of MARS. 

% UAVs are commonly equipped with inertial sensors, including accelerometers and gyroscopes, as well as non-inertial sensors such as the GPS and a compass, or other vision-based systems providing earth-frame position and heading information. To find an appropriate replacement for scenarios where inertial sensors are under attack, we adopt additional rotor speed sensors (i.e., tachometers) to provide indirect body-frame attitude information from the unique UAV physical~dynamics. 

% In our framework, we have 
MARS employs two sets of state estimators and control modes: \emph{standard} and \emph{resilient}. In the \emph{standard mode}, which provides optimal control performance when the UAV is not under attack, %MARS utilizes 
conventional UAV sensors %hubs for UAVs 
are used, including inertial accelerometers, gyroscopes, GPS, and compass. In the \emph{resilient mode}, MARS rejects IMU readings and incorporates additional rotor speed measurements, achieving strong attack resiliency with \emph{acceptable} control performance to continue mission flights. The switching between the modes is determined by a \emph{real-time anomaly detector} (AD) that monitors the estimation residuals between the %inertial sensor 
IMU measurements and the resilient state estimates. If %a system-level 
an anomaly is detected, the system switches~from the standard mode to the resilient mode under a \emph{multi-stage dynamic flight recovery strategy}, securing the UAV into near-hovering conditions %and then resumes 
before resuming~the~flight. 

% In what follows, we 
We now describe MARS in more detail. {We first introduce the model-based resilient state estimation (RSE), including the approach to estimate the UAV net thrust and torque from rotor speeds and earth-frame velocities; as well as fusing the high-fidelity thrust and torque estimates with other %robust 
sensor measurements %for the reconstruction 
to obtain robust UAV attitude %information 
estimates using an Extended Kalman Filter (EKF) (Sec.~\ref{subsec:control_input_estimate} and Fig.~\ref{fig:rse}).} We then introduce the %anomaly detection procedures 
AD that combines CUSUM statistics and sliding window detection (Sec.~\ref{subsec:ad_model}), followed by the MARS multi-stage dynamic flight recovery strategy (Sec.~\ref{subsec:recovery_model}).

\subsection{Model-based Resilient State Estimation}
\label{subsec:control_input_estimate}

% Determining the actually applied UAV control inputs %to the quadcopter system 
% is crucial when designing state estimators -- even for the same control commands these might differ with the platforms or battery-level changes.\todo{check; K: Add the example of PX4?} In a standard estimator %with available inertial sensor measurements, 
% that uses IMU data, such as the one in PX4 autopilot~\cite{PX4}, the control inputs are estimated from\todo{check} the measured angular velocities and accelerations that define the motion of the quadroter; %However, these readings become unreliable when inertial sensors are compromised, making the identification of an alternative source to capture the system control inputs necessary. 
% this is the reason why IMU attacks are so effective. On the other hand, the UAV control input $\mathbf{u}$ %to the quadcopter system 
% can also be described as the \emph{\textbf{collective}} thrust ($\mathbf{f}$) and torque ($\bm{\tau}$) applied at the quadrotor's center of mass -- i.e., $\mathbf{u}=\begin{bmatrix} \mathbf{f}^{\mathcal{B}}, \bm{\tau}^{\mathcal{B}}\end{bmatrix}^T$.

Determining the body-frame attitude %information, 
states, including the attitude quaternions and angular velocity, is crucial to perform real-time position and attitude control. In a standard UAV state estimator, the attitude information is acquired from IMU data, which are the measured angular velocities and accelerations that define the motion of the quadrotor. This is the reason why IMU attacks are so effective. On the other hand, we note that the attitude %information 
states can also be estimated from the \emph{\textbf{net}} thrust ($\mathbf{f}$) and torque ($\bm{\tau}$) applied at the quadrotor's center of mass; thus, we define $\mathbf{u}=\begin{bmatrix} \mathbf{f}^{\mathcal{B}}, \bm{\tau}^{\mathcal{B}}\end{bmatrix}^T$. We now %In this section, we 
%xdemonstrate %the process of acquiring 
show that accurate estimates of the \textit{net} torque and thrust can be obtained from the propeller rotational speed and earth-frame velocities \emph{under near-hovering assumptions}.
To achieve this we employ %s of 
the vehicle's dynamical model.

\subsubsection{{Near-hovering Thrust and Torque Estimation}}
\label{subsec:near_hovering_estimation}

We consider the force $\vec{F}_i$ and torque $\vec{M}_i$ generated by the $i^{th}$ propeller, applied on the UAV rotor's center of mass $A_i$, under near-hovering condition, which can be modeled as~\cite{martin2010true}:
%
% %
% \begin{equation}
% \begin{split}
%     \vec{F}_i &= \vec{F}_{rotor,i} + \vec{F}_{drag, i} + \vec{F}_{angular, i}\\
%     \vec{M}_i &= \vec{M}_{rotor,i} + \vec{M}_{drag, i} + \vec{M}_{rm, i} + \vec{M}_{angular, i},
% \end{split}
% \end{equation}
% %
%
\begin{equation}
\begin{split}
\vec{F}_i &= -a \omega_i^2 \vec{k}_b - \omega_i \left( \lambda_1 \vec{V}_{A_i}^{\perp} - \lambda_2 \vec{\Omega} \times \vec{k}_b \right)  \\ &\quad + \epsilon_i \omega_i \left( \lambda_3 \vec{V}_{A_i} \times \vec{k}_b - \lambda_4 \vec{\Omega}^{\perp} \right)  \\ 
\vec{M}_i &= -b \epsilon_i \omega_i^2 \vec{k}_b - \omega_i \left( \mu_1 \vec{V}_{A_i}^{\perp} + \mu_2 \vec{\Omega} \times \vec{k}_b \right)  \\
& \quad - \epsilon_i \omega_i \left( \mu_3 \vec{V}_{A_i} \times \vec{k}_b + \mu_4 \vec{\Omega}^{\perp} \right);
\end{split}
\label{eq_full_control_input}
\end{equation}
%
%where $\vec{F}_i$ and $\vec{M}_i$ denotes the force and torque vector applied at the center of mass of the $i$-th rotor. 
here, $\omega_i$ is the propeller rotational angular velocity with direction $\epsilon$ ($\epsilon >0$ represents counter clock-wise (CCW) revolution and %a negative one represents 
$\epsilon < 0$ clock-wise (CW) revolution); $\vec{\Omega}$ and $\vec{\Omega}^{\perp}$ denote the angular velocity of the quadrotor and its perpendicular projection on the rotor plane; unit vector $\vec{k}_b$ points down and is perpendicular to the rotor plane; $\vec{V}_{A_i}$ and $\vec{V}_{A_i}^{\perp}$ represent the earth-frame linear velocity and the its perpendicular projection on the rotor plane; and $a, b$, $\lambda_1, \lambda_2, \lambda_3, \lambda_4, \mu_1, \mu_2, \mu_3, \mu_4 $ are rotor's physical parameters. %constants. % of the rotor. 

%When the drone is hear-
While near-hovering, the term associated with body-frame %quadcopter 
angular velocity can be neglected, and %we consider 
the four rotors %share
have (almost) the same earth-frame velocities as the quadrotor center of mass. Thus,~\eqref{eq_full_control_input} can be simplified as:
\begin{equation}
\begin{split}
\vec{F}_i &\approx -a \omega_i^2 \vec{k}_b - \omega_i \lambda_1 \vec{V}_{A_i}^{\perp} +\epsilon_i \omega_i \lambda_3 \vec{V}_{A_i},\\
\vec{M}_i &\approx -b \epsilon_i \omega_i^2 \vec{k}_b - \omega_i \mu_1 \vec{V}_{A_i}^{\perp} - \epsilon_i \omega_i  \mu_3 \vec{V}_{A_i}.
\end{split}
\label{eq_rotor_torque_and_thrust}
\end{equation}

From~\eqref{eq_rotor_torque_and_thrust}, %indicates that 
under the near-hovering assumptions, the torque and thrust generated by an individual propeller are related only to the squared rotor speed and the earth-frame velocity of the UAV's center of mass. The net thrust and torque applied at the center of mass, % of the quadcopter, 
denoted by $C$, can then be approximated as the sum (over 4 propellers/rotors)
\begin{equation}
\mathbf{f}^{\mathcal{B}}  = \sum\nolimits_{i=1}^{4} \vec{F}_i, \quad
\bm{\tau}^{\mathcal{B}} = \sum\nolimits_{i=1}^{4} \vec{CA}_i \times \vec{F}_i + \vec{M}_i.
\label{eq_simplified_control_input}
\end{equation}

From~\eqref{eq_simplified_control_input}, the net torque and thrust of the quadrotor, which drive the drone dynamics as captured in~Sec.\ref{appendix:drone_model} (i.e.,\eqref{eq:model1} and~\eqref{eq:model2}) 
% Eq.indicates that  
can be computed using individual propeller torque and thrust, which rely \textit{only on rotor speed and earth-frame velocity measurements}. This approach enables UAV net thrust and torque estimation without onboard inertial sensors, which can be further tuned (Sec.~\ref{sec:tor_compensation}) to facilitate robust estimation of the attitude state (Sec.~\ref{subsec:rse_model}).

\subsubsection{Dynamic Torque Compensation}
\label{sec:tor_compensation}
\eqref{eq_simplified_control_input} provides a thrust and torque estimate under hovering conditions, using the rotor speed and earth-frame velocity measurements. Yet, drones are unlikely to remain in near-hovering conditions during various %autonomous 
flights, and attacks may occur in both hovering and non-hovering situations. Thus, when an attack is detected, it is critical to temporarily `stop' the drone, restore near-hovering, and then resume flight. To expedite the braking and restoration process and enhance the UAV's stability %of drone control 
during this procedure, a dynamic torque compensation related to the body-frame velocity, which indicates the direction of local drone movement, can be added to~\eqref{eq_simplified_control_input}. 

The compensated torque can be expressed as:
\begin{equation}
\bm{\tau}_{cp}^{\mathcal{B}} = \bm{\tau}^{\mathcal{B}} + K_{cp} \vec{V}_{C}^{\mathcal{B}} + \bm{\tau}_b^{\mathcal{B}},
\label{eq_compensated_control_input}
\end{equation}
%
 % $\vec{V}_{C}^{\mathcal{B}} = \mathbf{R}(\mathbf{q}) \vec{V}_{C}$ 
%
where $K_{cp} = \begin{bmatrix}
    k_{cp,x}, k_{cp,y}, k_{cp,z}
\end{bmatrix}^T$ is a set of tuned coefficients from experiments for the braking mechanism; $\vec{V}_{C}^{\mathcal{B}}$ is the projection of the earth frame linear velocity $\vec{V}_{C}$ onto the body frame. Here, we only consider the rotation along the Z-axis characterized by the measured yaw angle $\psi$ from the compass in this projection, as roll and pitch angles are not directly accessible. Also, $\bm{\tau}_b^{\mathcal{B}}$ is the torque bias term used to tune the net torques for optimal hovering stability, and $K_{cp}$ and $\bm{\tau}_b^{\mathcal{B}}$ can be experimentally tuned for best performance. 

The intuition behind this torque compensation for UAV `braking' %the drone 
is the coupling effect of the quadrotor's translational and rotational movements---when the UAV is moving at a certain velocity, it has a corresponding roll and pitch angle that enables it to move in that direction. Hence, to `stop' the drone, an opposite angular momentum is needed, which can be achieved by adding a torque vector in the correct direction. The faster the drone is moving, the larger the opposite angular momentum that needs to be~applied.

To fully estimate the net thrust and torque, we need to consider all contributing factors, including not only the rotation speed of the four rotors but also the effects of propeller angular acceleration and the quadrotor's body-frame angular velocity. The challenge lies in the absence of useful information from the IMUs when they are under attack. Hence, the estimate of the thrust and torque solely using propeller speeds and earth-frame velocities will inevitably deteriorate as the drone moves further away from near-hovering conditions and exhibits more acute dynamic behavior. %We consider this 
This is an intrinsic limitation when trustworthy accelerometers and gyroscopes are unavailable. Still, we demonstrate that even with this level of thrust and torque estimation, we can design a \emph{nonlinear resilient state estimator that provides accurate body-frame estimates}, enabling drone attitude controllers to stabilize the drone and prevent~%immediate 
crashes.

\subsubsection{EKF-based Attitude Estimation}
\label{subsec:rse_model}

Typically, IMUs directly provide angular velocity and acceleration information $\bm{\Omega}^{\mathcal{B}}=\begin{bmatrix}\Omega_x,\Omega_y,\Omega_z\end{bmatrix}^T$ and ${a}=\begin{bmatrix}a_x,a_y,a_z\end{bmatrix}^T$ for the drone controllers as the information is directly observable from the accelerometer and gyroscope sensors. However, when IMU's %are compromised and 
sensor readings cannot be trusted, it is %needed 
critical to estimate the missing attitude information (i.e., the attitude state -- both the attitude quaternion and body-frame angular velocity) from another robust source; we achieve this using the combination of dynamic net thrust and torque estimates, and the remaining unaffected sensors. %the remaining sensors, including both the quaternion and body-frame angular velocity. 
This is a nonlinear state estimation problem, and we adopt the EKF to achieve better accuracy than linear estimation models. 

\begin{figure}[!t]
\centering
\includegraphics[width=0.95\columnwidth]{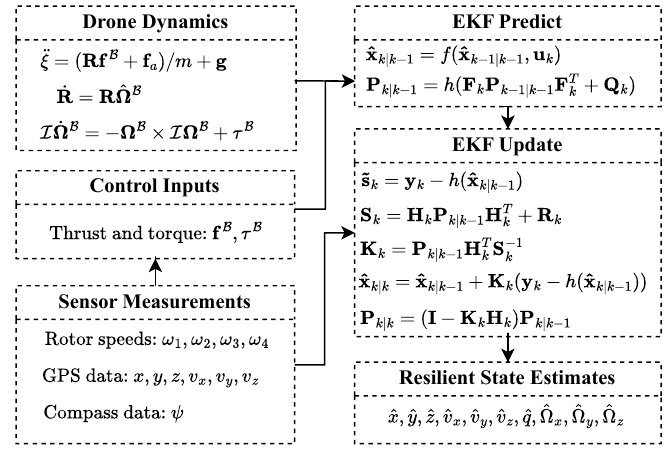}
\caption{Model-based resilient state estimator architecture.}
\label{fig:rse}
\end{figure}
% \todo{Luo: In Figure 3, shall we change Control Inputs to Control Estimates or Net Estimates?}

To capture the %quadcopter 
UAV's motion %of the quadcopter 
and enable necessary position and attitude control, we define the state variable $\mathbf{x}=\begin{bmatrix}x,y,z,v_x,v_y,v_z,q_w, q_x, q_y, q_z, \Omega_x,\Omega_y,\Omega_z\end{bmatrix}^T$, which includes the position, velocity, attitude and body-frame angular velocity. After IMU attacks, we rely on position and heading sensors, such as GPS and compass to provide \emph{noisy} observation data $\mathbf{y}=\begin{bmatrix}x,y,z,v_x,v_y,v_z,\psi\end{bmatrix}^T$. We also use tachometers to provide rotor speed measurements $\mathbf{\omega}=\begin{bmatrix}\omega_1, \omega_2, \omega_3, \omega_4\end{bmatrix}^T$, 
then used to estimate the net thrust and torque $\mathbf{u}=\begin{bmatrix} \mathbf{f}^{\mathcal{B}}, \bm{\tau}^{\mathcal{B}}\end{bmatrix}^T$, %that serve 
{as discussed in Sec.\ref{subsec:near_hovering_estimation}-\ref{sec:tor_compensation}.} % to the system: 

The intuition behind this estimation approach lies in replacing the IMU measured acceleration and angular velocity to thrust and torque from individual rotor speeds to perform body-frame attitude and angular velocity estimation, incorporating the UAV dynamics. Fig.~\ref{fig:rse} illustrates the process of generating resilient state estimates by incorporating sensor measurements with the drone's physical dynamics under the condition of compromised IMUs. A detailed formulation for this nonlinear EKF is provided in App.~\ref{appendix:ekf} -- note that we used the vector $\mathbf{u}=\begin{bmatrix} \mathbf{f}^{\mathcal{B}}, \bm{\tau}^{\mathcal{B}}\end{bmatrix}^T$ to denote the net thrust and torque estimates, as they provide inputs to the UAV dynamical model used in the EKF formulation.

\subsection{Statistical Anomaly Detection}
\label{subsec:ad_model}

To detect and respond to ongoing IMU attacks, % against the onboard IMUs, %inertial sensors, 
an efficient AD % anomaly detector (AD) 
is required. Statistical anomaly detection algorithms, including $\chi^2$, cumulative sum (CUSUM) and sequential probability ratio test (SPRT) have been deployed in robotic systems (e.g.,~\cite{jovanov_tac19,kwon2014analysis,quinonez2020savior}). Machine learning (ML)-based ADs have also been adopted for discovering embedded attack signatures from system logs (e.g.,~\cite{kravchik2018detecting}). 
In App.~\ref{appendix:ad_ht}, we provide a setup of anomaly detection as a hypothesis testing~problem.

\subsubsection{MARS-based CUSUM detection}

To practically solve the hypothesis testing problem, %we need to design a proper mapping for $D(Y_k)\to \{0, 1\}$. Therefore, 
we focus on one outstanding statistics, the \textit{MARS resilient estimate residual}, as the anomaly score to `evaluate' the level of anomaly. % and fulfill the mapping. 
This anomaly score represents how much the observed data deviates from the estimation, and can raise an alarm if that deviation exceeds some manually set threshold. In our design, we adopt CUSUM statistics on the estimate produced by the MARS resilient state estimator. 

Specifically, we continuously monitor accelerometer and gyroscope measurements and compute the residual:
\begin{equation}
    r_{k} = Y_{s,k} - h_{r}(\hat{x}_{r,k}),
\end{equation}
where $r_k$ is the residual between current resilient estimate $\hat{x}_{r,k}$ and current inertial sensor observation data $Y_{s,k}$, computed through the observation function $h_{r}$. Then the CUSUM statistics describing this difference is: %can be expressed as:
%
% \begin{equation}
$    S_{k} = max(0, S_{k-1} + r_{k-1} - b),
$
% \end{equation}
%
where $S_{k}$ and $S_{k-1}$ are the current and previous step CUSUM statistics; $r_{k-1}$ is the previous step estimate residual and $b$ is a constant to control the accumulation~of~error. 

If CUSUM statistics is larger than a pre-defined threshold $\lambda$, then a point anomaly $\alpha_{k}$ is raised, and the current CUSUM statistics is set back to zero ($S_{k}=0$); otherwise, this point is non-anomaly:
%
% \begin{equation}
$
    \alpha_{k} = 
    \begin{cases} 
    1, & \text{if } S_{k} > \lambda, \\
    0, & \text{otherwise}.
    \end{cases}
$
% \end{equation}
%

% The main novelty %contribution 
% of MARS anomaly detection lies in the use of the MARS resilient estimate residual, rather than relying on a specific statistical algorithm. Hence, besides CUSUM, other algorithms that track historical anomalies can be tuned to work~with~MARS.

\subsubsection{Sliding window in online detection}

While CUSUM statistics reflect the accumulation of error over time, simply relying on point anomalies alone increases the possibility of false positives especially when the drone experiences sudden disturbances. Therefore, in online detection scenarios, we adopt a sliding-window technique to further analyze the CUSUM statistics and build a more robust anomaly decision-making process. The status of a system anomaly is determined based on the ratio of CUSUM anomalies within a certain time window. By grouping the point anomalies into a sliding window with length $l$ and defining a system anomaly indicator, the current detection rate $DR_k$ is %computed as:
\begin{equation}
\begin{split}
     DR_{k} = \frac{1}{l} \sum\nolimits_{i=0}^{l} \alpha_{k-l+i+1}.
\end{split}
\end{equation}

Based on this detection rate, we can check %whether or not 
if the UAV is under attack and if further action needs to be taken: % accordingly:
\begin{equation}
    \alpha_{s, k} = 
    \begin{cases} 
    \text{True}, & \text{if } DR_{k} > p, \\
    \text{False}, & \text{otherwise},
    \end{cases}
\end{equation}
where $p$ refers to the false positive rate that the system tolerates without giving an alarm. Since the existing attacks on inertial sensors are generally destructive with significant impact at time of attack, we choose a relatively small $p=0.005$. Defining the starting time of the %anomaly detector 
AD at $k=0$, 
a detailed description of our detection and recovery process is shown in Alg.~\ref{alg:mars_ad}, %illustrated in the following algorithm:
where $Y_{s,k}$ and $Y_{r,k}$ are sensor measurements in the standard and resilient control modes -- i.e., $Y_{s,k}$ contains IMU %accelerometer and gyroscope 
readings and other sensor measurements, while $Y_{r,k}$ excludes IMU readings but includes %additional 
rotor speed measurements.

\begin{algorithm} [!t]
\caption{MARS Anomaly Detection}
\label{alg:mars_ad}
\begin{algorithmic}
    % \STATE MARS Design
   \STATE  \textbf{Input:} $Y_{s,k}$ for sensor readings in standard control mode, and  $Y_{r,k}$ for sensor reading in resilient control mode.
   \STATE \textbf{Output:} System anomaly $\alpha_{s, k} \in \{True, False\}$.
   \STATE \textbf{Initialization:} $k=0, S_{-1}=0, r_{-1}=0$.
  \WHILE{system is running}
   \STATE Update current sensor readings $Y_{s,k}$ and $Y_{r,k}$ 
  \STATE Standard estimate $\hat{x}_{s,k}=f_s(Y_{s,k})$
   \STATE Resilient estimate $\hat{x}_{r,k}=f_r(Y_{r,k})$
  \STATE Estimate residual $r_{k} = Y_{s,k} - h_{r}(\hat{x}_{r,k})$
   \STATE CUSUM statistics $S_{k} = max(0, S_{k-1} + r_{k-1} - b)$
  \IF{$S_{k}  > \lambda$}
  \STATE CUSUM point anomaly $\alpha_{k}=1$ \\ $S_{k} = 0$
   \ELSE 
  \STATE CUSUM point anomaly $\alpha_{k}=0$
   \ENDIF
  \STATE Detection rate $DR_{k} = \frac{1}{l} \sum_{i=0}^{l} \alpha_{k-l+i+1}$
   \IF{$DR_{k}  > p$}
   \STATE System anomaly $\alpha_{s, k}=\text{True}$. System under attack, $\hat{x}_{r,k}$ is sent to drone controllers.
   \ELSE  
  \STATE System anomaly $\alpha_{s, k}=\text{False}$. No attack detected, $\hat{x}_{s,k}$ is sent to drone controllers.
  \ENDIF
  \ENDWHILE
  \label{mars_algorithm}
\end{algorithmic}
\end{algorithm}

% \subsubsection{Sliding window trade-off} 
Adding a sliding window on top of CUSUM anomalies inevitably increases the %detector 
AD response time if an actual attack occurs. However, this window reduces the likelihood of false switches from the standard control to the resilient mode in the context of online detection and recovery. Later, in our evaluations, we demonstrate that this trade-off does not affect the recovery capability of MARS.

{The main novelty of MARS anomaly detection lies in the use of the MARS resilient estimate residual, rather than relying on a specific statistical algorithm. Hence, besides CUSUM, other algorithms that track historical anomalies can be tuned to work~with~MARS. MARS anomaly detector outperforms existing benchmarks in both detection rates and time-to-detection statistics, discussed in Section~\ref{subsec:ad_analysis}.}

\subsection{Multi-Stage Dynamical Flight Recovery}
\label{subsec:recovery_model}

While stable hovering ensures the capability of surviving inertial sensor attacks, it is highly likely for UAVs to encounter attacks during more dynamic mission segments. Thus, it is necessary to analyze the potential for recovering during e.g., aggressive maneuvers, 
beyond just hovering.  However, as is discussed in Sec.~\ref{subsec:control_input_estimate}, when a UAV is moving at a high speed, it deviates from near-hovering conditions, increasing the error in the thrust and torque estimates. This means that at the time of an attack, if the drone has a large pitch or roll angle, the torque and thrust estimates would suffer, providing less accurate attitude information. %to the drone controllers.

\begin{figure}[!t]
\centering
\includegraphics[width=\columnwidth]{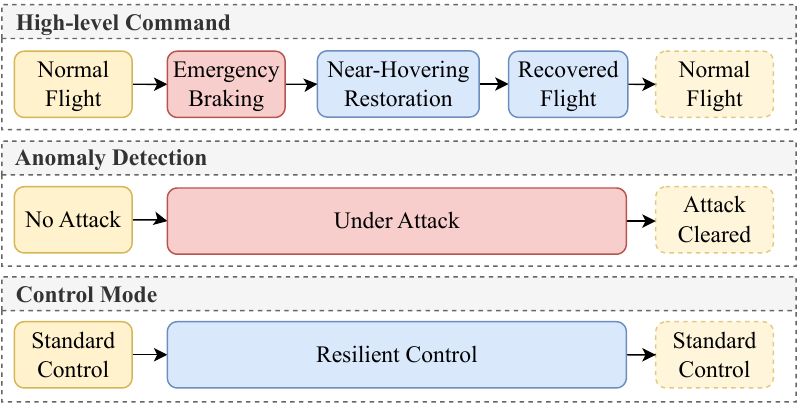}
\caption{Multi-stage dynamical recovery for arbitrary autonomous missions under the coordination of high-level command, AD status and control mode selection.}
\label{fig:recover_mission}
\end{figure}

Thus, we need to design a multi-stage dynamical recovery strategy to handle attack regardless of the %mission 
maneuver the UAV is performing during the attack. First, a `braking' stage is necessary for the UAV to restore near-hovering conditions (as discussed in Sec.~\ref{sec:tor_compensation}), allowing it to stabilize using accurate resilient state estimates. After recovering from the impact of the attacks and achieving a stable hovering state, the UAV %can then 
resumes its mission under resilient control until the threats are cleared. 
This is summarized 
% As shown 
in Fig.~\ref{fig:recover_mission}; the high-level controller (i.e., command) sends an immediate braking command once the onboard %anomaly detector 
AD detects an anomaly. Simultaneously, the controller switches to the resilient mode to begin the recovery process. The UAV then enters the near-hovering restoration stage, where the drone's control objective is set to hovering, in order to enable obtaining accurate resilient control estimates and optimal resilient control performance. Once the UAV achieves stable hovering, it can resume its %autonomous 
mission using the resilient control, entering the next stage: \emph{recovered flight} (using the estimator from Sec.~\ref{subsec:rse_model}). 
Finally, if the attack stops %is cleared (i.e., attack is no longer present in AD), 
(i.e., AD flag clears), the UAV switches back to normal flight, fully using its~IMU.

%% file: Evaluation.tex
\section{Attack Detection and Recovery Analysis}
\label{sec:eval1}

\subsection{MARS Implementation}

\begin{figure*}[!t]
\centering
\includegraphics[width=0.968\textwidth]{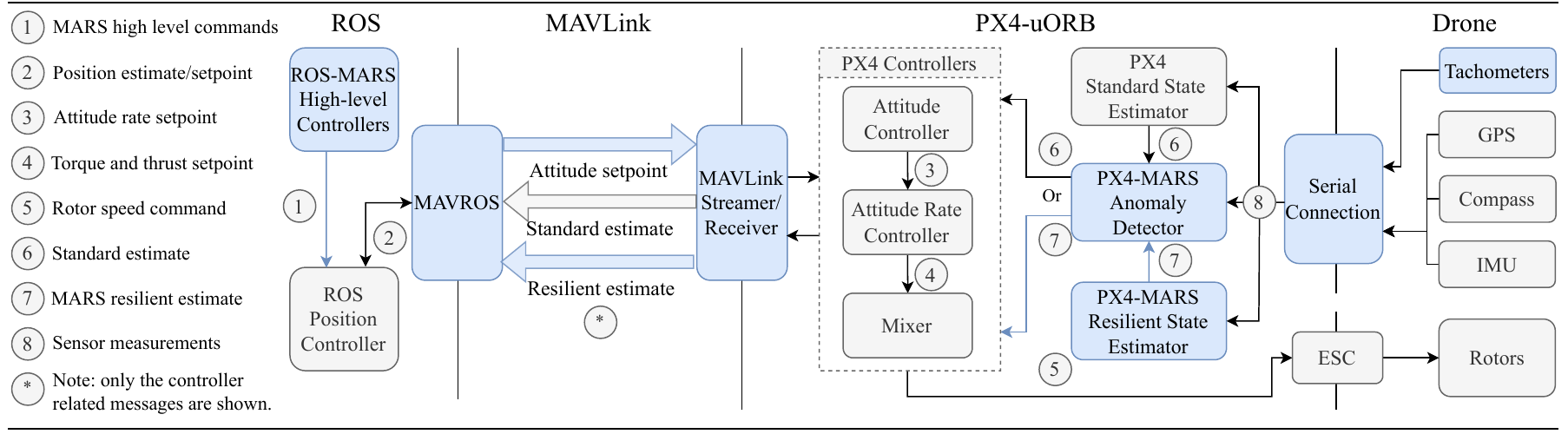}
\caption{MARS-PX4 autopilot architecture. Gray blocks: PX4 default modules; blue blocks: our customized or specially tailored PX4 modules and communication bridges for MARS compatibility in the PX4 autopilot ecosystem.}
\label{fig:mars_px4}
\end{figure*}

We implemented MARS % recovery 
 in the open-source autopilot PX4~\cite{7140074, PX4}; a customized MARS-PX4 autopilot with MARS modules was integrated into the PX4 uORB messaging environment and incorporated Robot Operating System (ROS) and MAVLink communication protocols for data exchange (Fig.~\ref{fig:mars_px4}). 
For evaluation, % effectiveness, %of our proposed method, 
we also implemented realistic inertial sensor attacks following the approaches from~\cite{son2015rocking, tu2018injected, jang2023paralyzing}. 
For simulation-based evaluations, we %also utilized
used %the software toolbox of an 
the open-source vehicle simulation platform %called
Prometheus~\cite{Prometheus}, powered by %Robot Operating System (ROS) 
ROS and the Gazebo robot simulator. %Using these tools, 

The ROS component consists of MARS high-level controllers, where the MARS recovery module and the position controller are instantiated, interacting with MAVLink communication through the MAVROS package. MAVLink messaging interconnects ROS high-level communication with PX4 uORB messaging via MAVLink streaming and receiving mechanisms, which are specially customized to transmit resilient state estimates. The core flight controller modules inside PX4 receive control commands from the high-level controllers, compute the desired rotor speeds, and send them to the Electronic Speed Control (ESC) units. The state estimates required by the controllers can be provided either by the PX4 standard state estimator or our MARS Resilient State Estimator (MARS-RSE), as determined by the MARS Anomaly Detector (MARS-AD). 

We first conducted %our attack detection and recovery 
MARS analysis in Software-in-the-Loop (SITL) environments, %making it possible to migrate to real-world drones equipped with 
`connecting' Gazebo simulated drones controlled by our customized MARS-PX4 autopilot. % without the need for major modifications. 
Tables \ref{tab:px4_sitl_sensors} and %Table~
\ref{tab:px4_sitl_controllers} in App.~\ref{appendix:system_params} list the virtual sensors, estimators, and controller specifications for PX4~SITL~evaluations. 

Using this setup, we initially evaluated the accuracy of MARS resilient state estimation by computing the thrust and torque and angular information estimate error %with simulator ground truth 
(Sec.~\ref{subsec:rse_analysis}). 
Then, for four different inertial sensor attacks, % and test them with 
we evaluated the effectiveness of MARS-AD against benchmark detectors %by monitoring 
% in terms of the average detection response time and the rotor speed fluctuations caused by pre-detection attacks 
(Sec.~\ref{subsec:ad_analysis}).
Moreover, we evaluated the complete MARS framework on drone hovering missions and compare its performance to three representative inertial sensor recovery approaches from~\cite{jeong2023rocking, tu2019flight}. We demonstrated the superiority of MARS over existing methods in survival time and control smoothness (Sec.~\ref{subsec:hover_analysis}). 
Finally, we evaluated the MARS multi-stage dynamical recovery strategy in dynamic tracking missions (Sec.~\ref{subsec:move_analysis}), showing MARS capability to complete flight missions even %in the presence of 
under attack, 
with only small performance~degradation.
% and analyze the resulting tracking error and mission completion time.

\subsection{Resilient State Estimate Effectiveness}
\label{subsec:rse_analysis}

The MARS resilient state estimator is designed to provide accurate state estimates for drone controllers, to ensure stable flights even under attack. The sensor group that feeds into the MARS resilient state estimator consists of position, heading, and rotor speed sensors. The key to MARS resilient state estimation is leveraging tachometer rotor speed measurements to calculate quadrotor thrust and torque, which are then used to generate body-frame attitude and angular velocity state estimates using the EKF. These estimates must meet the requirements of the drone attitude controller in terms of \emph{high accuracy} and \emph{proper update frequency}. % in tune with the controller itself.

\begin{figure}[!t]
\centerline{\includegraphics[width=\columnwidth]{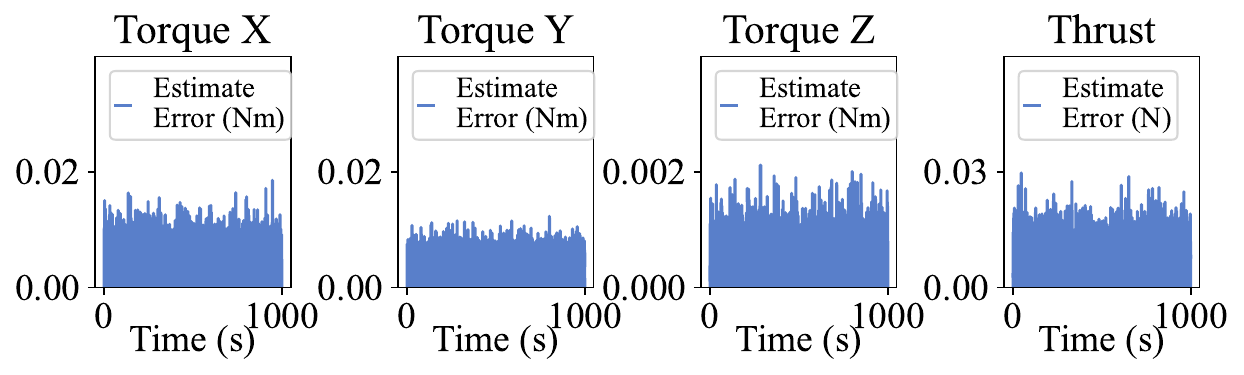}}
\caption{FRD frame thrust and torque estimation errors. }
\label{fig:control_input_estimate_error}
\end{figure}

To verify the accuracy, % of our control input estimates, 
we conducted sample runs in %a simulated 
the SITL environment %where we had access to 
with virtual tachometers providing rotor speed measurements. %, enabling the torque and thrust to be updated at the frequency of the attitude controller. 
Comparing with the ground truth torques and thrust applied at the center of gravity of the simulated quadcopter, Fig.~\ref{fig:control_input_estimate_error} shows torque and thrust estimate error in a $1000~s$ hovering sample run; %. It indicates 
the X-Y axis torque error stays below 0.02 $Nm$, the Z axis below 0.002 $Nm$, and the thrust below 0.03 $N$, way within the required levels.

\begin{figure}[t!]
    \centering
    \subfloat
    {\includegraphics[width=0.50\columnwidth]{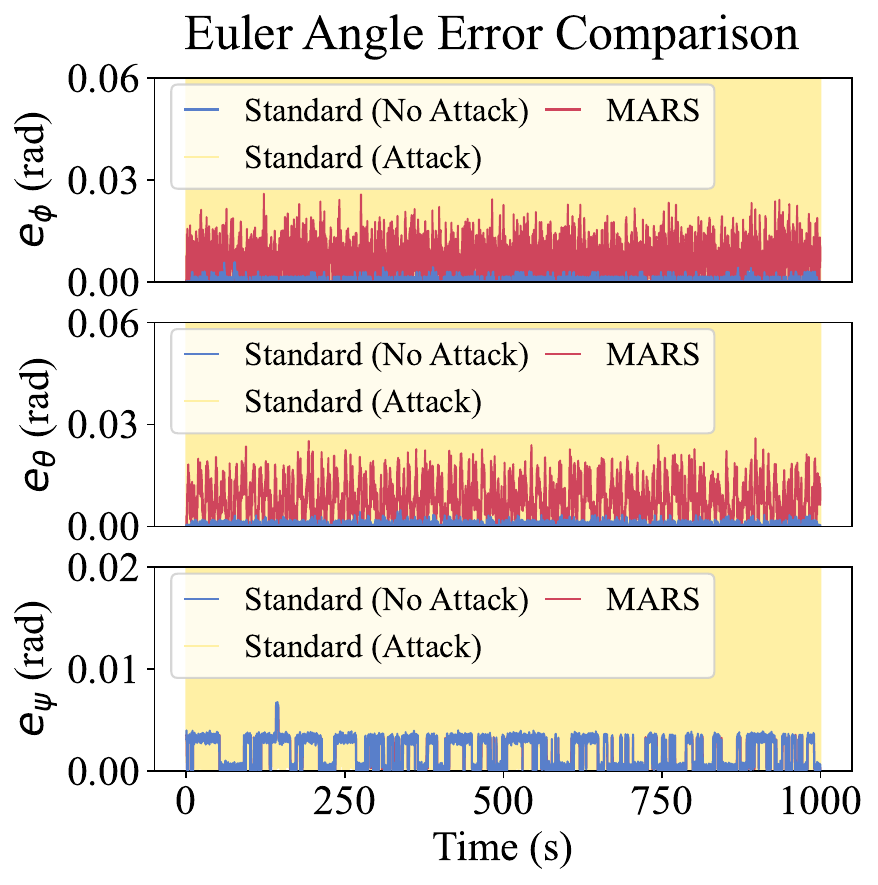}}
    \hfil
    \subfloat{\includegraphics[width=0.50\columnwidth]{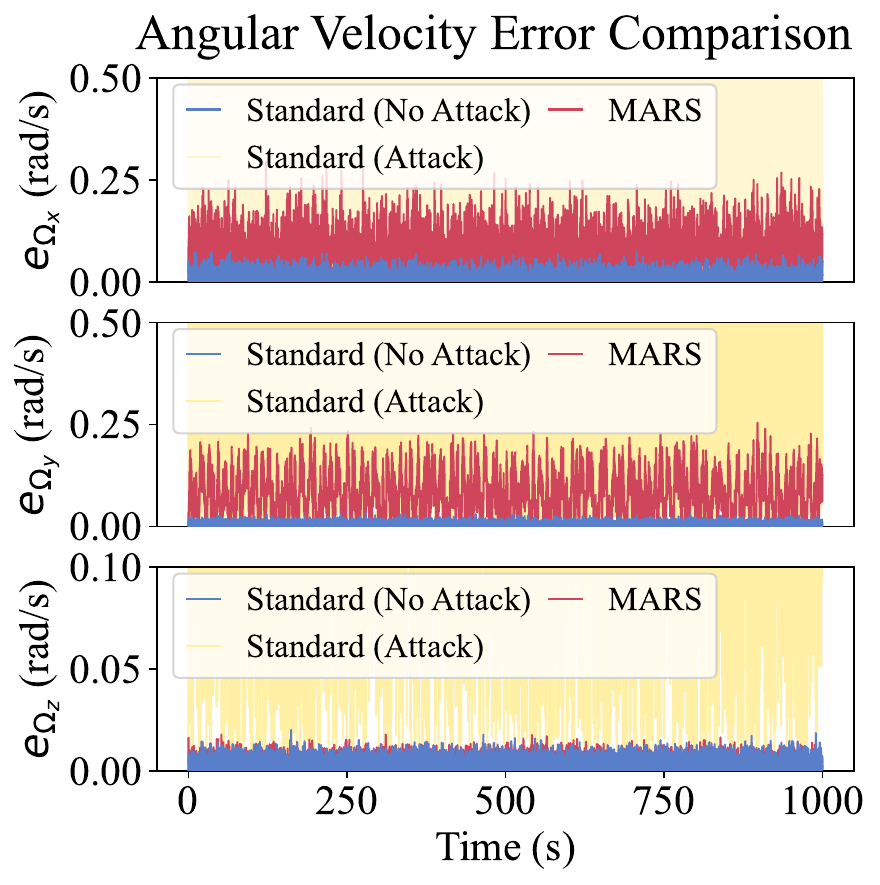}}
    \caption{Attitude and angular velocity estimate error comparison between MARS resilient state estimator and standard estimator. Left: Euler angle; right: angular velocity.}
    \label{fig:angular_estimate_error}
\end{figure}

With stable and accurate estimates for the applied % access to accurate 
thrust and torque, Fig.~\ref{fig:angular_estimate_error} shows that MARS resilient state estimator produces high-fidelity attitude and body-frame angular velocity data, 
% by incorporating the physical %invariant 
% model of the quadrotor. This approach enables 
enabling the drone to swiftly recover attitude control %that has been 
disrupted due to attacks on inertial sensors. Specifically, Fig.~\ref{fig:angular_estimate_error} compares the estimation error for three different scenarios: standard estimator without attack (blue), standard estimator under attack (yellow line), and MARS-RSE regardless of attack (red). While the standard estimator achieves optimal accuracy without attacks, due to the use of IMUs, it is highly vulnerable to IMU attacks (note the levels of the yellow line). In comparison, MARS-RSE achieves %near-optimal estimate effectiveness 
low error levels that allow for flight stabilization while being \emph{robust to IMU attacks}. % on inertial sensors}. 
The Z-axis estimates have much lower error, both in angle and angular velocity, as the yaw angle is directly observable from the compass sensor. In Sec.~\ref{subsec:hover_analysis}, we show that these level of accuracy is sufficient for reconstructing stable attitude control under IMU attacks. 
% in the presence of IMU sensor corruptions. 

\subsection{Detection of Inertial Sensor Attacks}
\label{subsec:ad_analysis}

MARS-RSE is only effective when IMU attacks %on inertial sensors can be 
are quickly and accurately detected, prompting a timely switch to resilient control. % to prevent danger.
Thus, we also evaluated the MARS-AD (i.e., accuracy and time-to-detection), focusing on both the reported and potentially more subtle IMU attacks.

\subsubsection{Design of Inertial Sensor Attacks}

We consider two categories of feasible and destructive inertial sensor attacks: acoustic resonant attacks~\cite{son2015rocking, tu2018injected} and %electromagnetic interference 
EMI attacks~\cite{jang2023paralyzing}. For %acoustic resonant attacks, 
the former, we examine three different attack variations: Denial of Service (DoS), Side-Swing, and Switch attacks. The DoS attack involves un-modulated resonance injection~\cite{tu2018injected}; 
whereas Side-Swing and Switch attacks are modulated versions of the DoS attack through attack manipulation. %; note that these two they have not been previously reported.\todo{check} 
Intuitively, the Side-Swing attack retains only the positive or negative components of the attack signal, setting the unwanted half to zero. The Switch attack  %, on the other hand, 
flips the unwanted half to the opposite direction, resulting in entirely positive or negative attack values. All three resonant attacks account for the sampling drift effect, which refers to the imprecise sampling intervals of the sensors, causing the attack signal to disperse over the bandwidth. To evaluate %our method 
MARS under the worst-case scenarios, we inject resonant signals into all three sensing axes of both accelerometers and gyroscopes simultaneously, achieving the maximum possible oscillation.

Unlike acoustic resonant attacks, the effects of %electromagnetic interference (EMI) 
EMI attacks are more unpredictable, potentially causing sensor saturation, inaccurate values from unexpected data packets, and persistent data loss~\cite{jang2023paralyzing}. To maximize the attack impact, we focus on the representative case of sensor saturation for EMI attacks, injecting signals at the sensing limits into both accelerometers and gyroscopes. 
More details about attack design are provided in 
%More detailed configurations for implementation are shown in 
App.~\ref{appendix:sitl_attack_profile} (e.g., Fig.~\ref{fig:attack_profiles} and~Table~\ref{tab:px4_attack_vectors}).

% In the simulated environment, we use the IMU sensor profile from the ICM-456xy MEMS Motion Sensor~\cite{ICM456XY}, with a full detection range for the accelerometer and gyroscope of $\pm~4000$ $dps$ and $\pm~32$ $g$, which correspond to $\pm~300 ~{m/s}^2$ and $\pm~70~rad/s$, respectively. And we set the normal range for accelerometer and gyroscope readings at $\pm~100~{m/s}^2$ and $\pm~10~rad/s$. As a result, we set the amplitude of the acoustic resonant attack at the maximum value within the normal detection range, and the amplitude of the electromagnetic interference saturation attack at the maximum value of the full range. The resonant frequency of drone onboard MEMS devices typically ranges from $20~kHz$ to $30~kHz$, while our system sampling frequency is $250~Hz$. After the signal aliasing effect, the induced frequency lies in the range of $0$ to $125~Hz$. To select an appropriate acoustic attack induced frequency, we chose $100~Hz$ and introduced a sampling drift of $500$ $\mu$s to simulate attack signal dispersion in real-world sensors. The attack vectors are illustrated in Fig.~\ref{fig:attack_profiles}. More detailed configurations for implementation are shown in Table~\ref{tab:px4_attack_vectors} in App.~\ref{appendix:sitl_attack_profile}.

\subsubsection{Accuracy and Time-to-Detection for MARS-AD}

To evaluate the MARS-AD based on the MARS-RSE residual and the CUSUM statistical test, we compare it with two benchmark %anomaly detectors
ADs: standard $\chi^2$ and standard CUSUM. Both benchmarks utilize the Mahalanobis distance produced by an IMU-based standard state estimator as their decision statistic but employ different statistical algorithms. All ADs employ the same sliding window approach to determine the system anomaly status from point anomalies. %To comprehensively compare them, 
For thorough comparison, we considered $10$ datasets for each of the four attack profiles, where each dataset consists of a no-attack window and an attack window of the same length ($20~s$), resulting in a total of $40$ datasets $\times$ $20~s$ each $\times$ $250~Hz$ sampling rate $=$ $20,000$ data points. Moreover, the maximum attack power is used as the worst-case scenario aiming at drone crashes. %in which an attacker attempts to bring down the drone.

\begin{figure}[!t]
\centering
\includegraphics[width=\columnwidth]{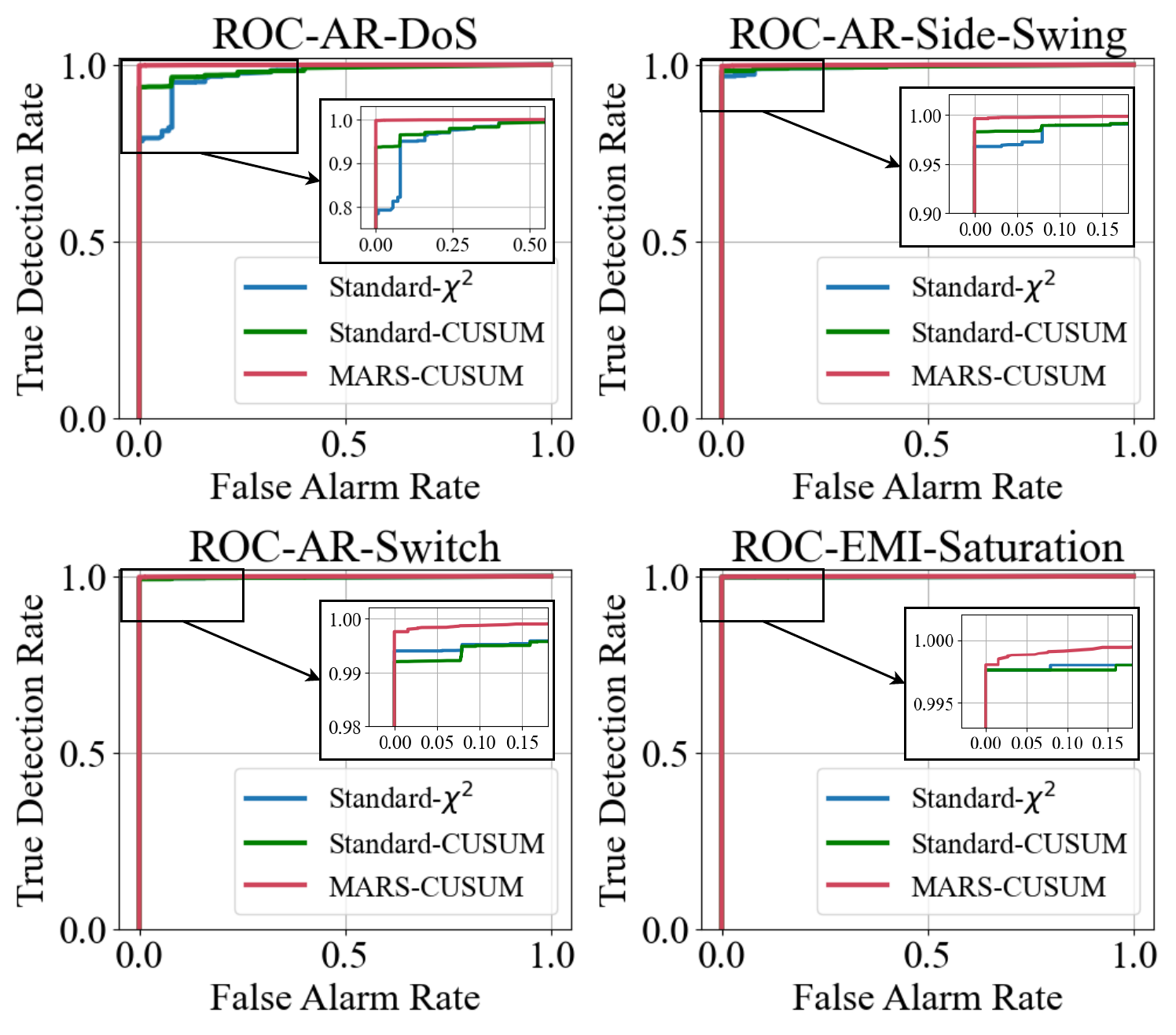}
\caption{Anomaly detectors ROC curves.}
\label{fig:anomaly_detection_roc}
\end{figure}

Fig.~\ref{fig:anomaly_detection_roc} illustrates the detection accuracy of the three detectors using Receiver Operating Characteristic (ROC) curves; 
% , and Table~\ref{tab:offline_anomaly_detection_accuracy} further compares the true detection rates of the ADs when the false alarm rate is set at 0.01. 
e.g., when the false alarm rate is set at 0.01, for AR-DoS attacks the true detection rates of $\chi^2$, Standard CUSUM and MARS CUSUM are 0.792,
0.937, and 0.998 respectively.
The results show that the MARS-AD outperforms the two standard estimator-based detectors by achieving higher true detection rates and lower false alarm rates.

% \begin{table}[!t]
%     % \captionsetup{skip=10pt}
%     \caption{AD true detection rate when $P_{FA}=0.01$.}
%     \centering
%     \renewcommand{\arraystretch}{1.3}
%     \begin{adjustbox}{width=\columnwidth}
%     \begin{tabular}{c c c c c}
%     % \hhline{=====}
%     \hhline{-----}
%     \multirow{2}{*}{\makecell[c]{\textbf{Anomaly}\\\textbf{Detectors}}} & \multicolumn{4}{c}{\textbf{Anomaly Detector True Detection Rate} ($P_{FA}=0.01$)} \\ 
%     \cline{2-5}
%     \cline{2-5}
%      & \makecell[c]{AR-DoS} & \makecell[c]{AR-Side-Swing} & \makecell[c]{AR-Switch} & \makecell[c]{EMI-Saturation} \\
%     \hhline{-----}
%     Standard $\chi^2$ & 0.792 & 0.968 & 0.994 & 0.998  \\ \hline
%     Standard CUSUM & 0.937 & 0.982 & 0.992 & 0.998  \\ \hline
%     MARS CUSUM & 0.998 & 0.996 & 0.998 & 0.998  \\ \hline
%     % \hhline{=====}
%     \hhline{-----}
%     \end{tabular}
%     \end{adjustbox}
%     \label{tab:offline_anomaly_detection_accuracy}
% \end{table}

\begin{table}[!t]
    % \captionsetup{skip=10pt}
    \caption{Anomaly detector average response time.}
    \centering
    \renewcommand{\arraystretch}{1.3}
    \begin{adjustbox}{width=\columnwidth}
    \begin{tabular}{c c c c c}
    % \hhline{=====}
    \hhline{-----}
    \multirow{2}{*}{\makecell[c]{\textbf{Anomaly}\\\textbf{Detectors}}} & \multicolumn{4}{c}{\textbf{Anomaly Detector Average Response Time ($s$)}} \\ 
    \cline{2-5}
    \cline{2-5}
     & \makecell[c]{AR-DoS} & \makecell[c]{AR-Side-Swing} & \makecell[c]{AR-Switch} & \makecell[c]{EMI-Saturation} \\
    \hhline{-----}
    Standard $\chi^2$ & 1.512 & 0.323 & 0.060 & 0.024  \\ \hline
    Standard CUSUM & 0.634 & 0.176 & 0.079 & 0.024  \\ \hline
    MARS CUSUM & 0.024 & 0.039 & 0.024 & 0.020  \\ \hline
    % \hhline{=====}
    \hhline{-----}
    \end{tabular}
    \end{adjustbox}
    \label{tab:offline_anomaly_detection_times}
\end{table}

In addition to accuracy, timely attack detection is critical for the success of subsequent real-time mitigation and recovery measures.
Table~\ref{tab:offline_anomaly_detection_times} shows the average AD response times for different attacks. Again, our MARS-AD %detector 
achieves low detection times across all attack profiles, while the standard estimator-based detectors take longer to identify anomalies and struggle with less acute attack vectors.

Further, since the reported %inertial sensor 
IMU attacks have overwhelmingly powerful effects, 
% it is not %an outstanding challenge 
% very challenging for %anomaly detectors 
ADs detect them with reasonably good accuracy and response times (in Fig.~\ref{fig:anomaly_detection_roc}, all 3 ADs have decent accuracy). However, it is highly likely that more sophisticated attack variants could be developed in cyber attack paths, or future physical IMU attacks might have capabilities  to compromise the UAV system in a more stealthy manner. %To address this possibility, 
Thus, we also conducted tests with two more modulated attacks for the four attack profiles: \emph{Step Amplitude} and \emph{Ramp Amplitude}. The detailed results shown in App.~\ref{appendix:anomaly_detection}, further demonstrate that the MARS-AD significantly outperforms the benchmark detectors %in detecting 
even with these more stealthy attacks. 

\subsection{Hovering Stabilization under Attack}
\label{subsec:hover_analysis}

Once an anomaly is detected, the system %should 
switches to the resilient control mode to avoid immediate crashes. MARS involves reconstructing body-frame attitude information through a robust sensor fusion that incorporates position, heading, and propeller revolutions. 

% \subsubsection{Benchmark recovery methods} 
We compared MARS with three existing recovery methods:

\vspace{2pt}
\noindent\textbf{Low-pass filtering (LPF)}: LPF is one of the most commonly used heuristic filters in autopilots for its simplicity and effective noise reduction (e.g.,~\cite{ban2013integral}). We adopted the PX4 built-in low-pass filter functions with a $30~Hz$ cutoff frequency.
% for both the accelerometer and gyroscope.

\vspace{2pt}
\noindent\textbf{Deep-Auto-Encoder (DAE)~\cite{jeong2023rocking}} (App.~\ref{appendix:recovery_methods} provides the training details). 
To ensure that the DAE model meets the $250~Hz$ update frequency required for attitude controllers, which was highlighted as one of the limitations in~\cite{jeong2023rocking}, we slowed down simulation by setting the real-time factor to $0.2$ (i.e., 5x slower than real-time) and monitored the ROS data flow to confirm an actual $4~ms$ model inference time.

% DAE represents the emerging machine learning-based filtering techniques that excel at capturing embedded patterns in attack profiles compared to heuristic filters. For training purposes, we recorded datasets of drone hovering and waypoint-visiting missions with randomly generated reference waypoints and headings. We then built our attack recovery dataset by adding three different variations of acoustic resonant attacks, resulting in a total of $548,000$ data samples in our training set. We did not include the EMI attack profile in the training set, as it is trivial for filtering-based methods to learn from saturated sensor measurements. Using the attack-recovery dataset, we trained two deep autoencoder models for the accelerometer and gyroscope, achieving Mean Square Error (MSE) test losses of $0.046$ and $0.005$, respectively. Additionally, to ensure that our DAE model meets the $250~Hz$ update frequency required for attitude controllers, we slowed down our simulation by setting the real-time factor to $0.2$ (which is five times slower than real-time) and monitored the ROS data flow to confirm an actual $4~ ms$ model inference time.

\vspace{2pt}
\noindent\textbf{Complementary Attitude Feedback (CAF)}: CAF is an attitude reconstruction method using position information and geometric relationship~\cite{tu2019flight}. It relies on position and heading data to geometrically calculate the Euler angles and feed them into drone controllers. Therefore, the update frequency of CAF is tied to the position sensor, not with the attitude controller, which is unrealistic in practice. %However, in our implementation
For evaluation, we maintain a $250~Hz$ output frequency but only update the estimate at the GPS frequency.

% \noindent\textbf{Evaluation Results}
\subsubsection{Evaluation Results}
\begin{figure}[!t]
\centering
\includegraphics[width=\columnwidth]{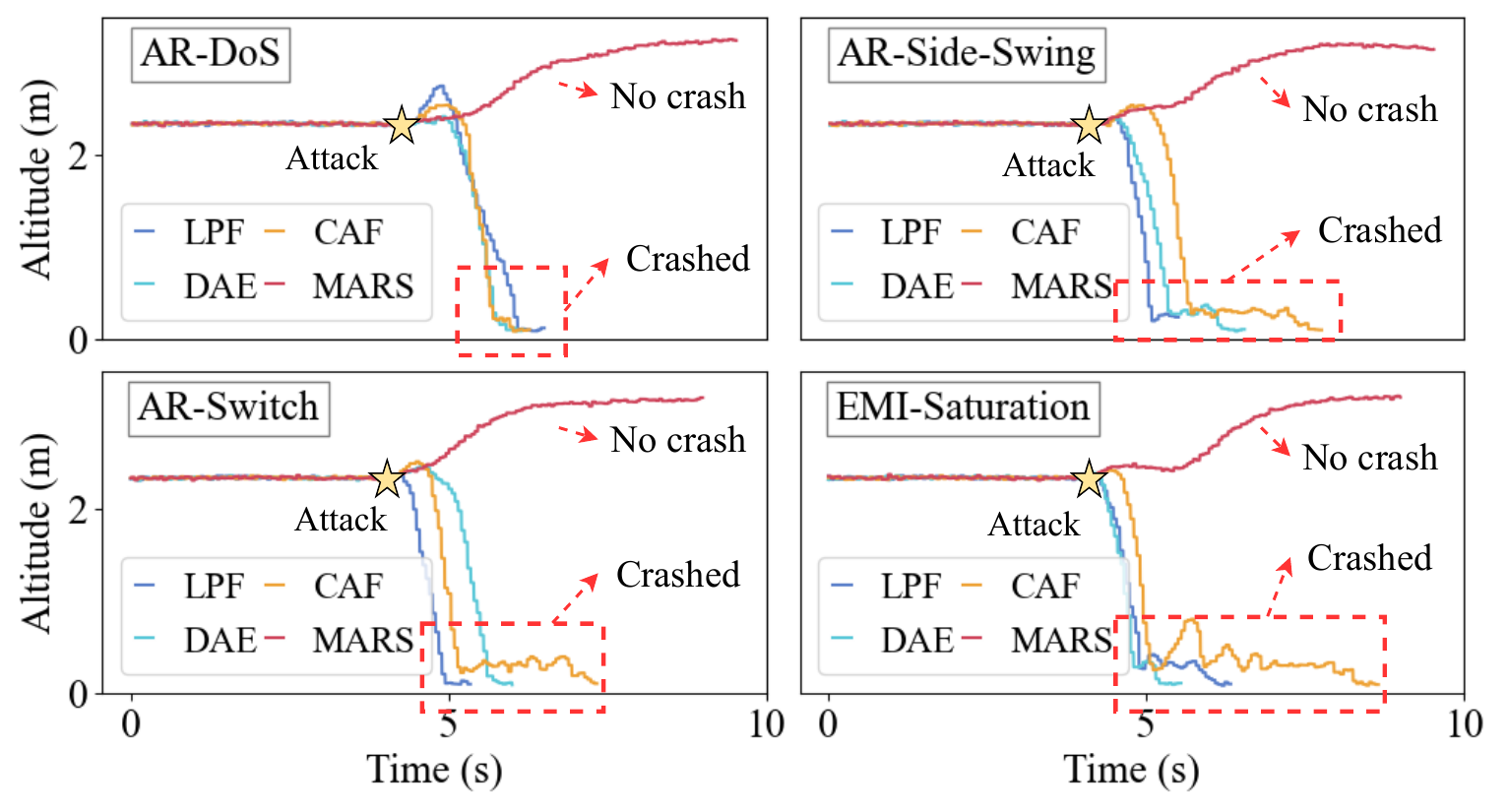}
\caption{The UAV altitude in hovering missions under~four inertial sensor attacks with different recovery frameworks.}
\label{fig:hovering_altitude_comparison}
\end{figure}

We first tuned all methods to ensure their estimates can \textit{control the drone stably in attack-free scenarios}. Hence, any emerging drone crash does not result from poor control design but from the unmitigated impact of %inertial sensor 
the IMU attacks. %We deployed them in PX4 autopilot and tested them 
Evaluations we done in SITL hovering missions %with attacks. In this analysis, the quadcopter is set to take off 
with the UAV taking off and hovering at a given location when %simulated 
the IMU attacks start. Fig.~\ref{fig:hovering_altitude_comparison} tracks the altitude of the quadrotor in the the hovering missions under the four recovery frameworks when four attack vectors were injected. The three representative methods failed to maintain the drone's stability, causing it to fall to the ground within $2~s$, while MARS made the UAV survive the attacks.

\begin{figure}[!t]
\centering
\includegraphics[width=0.94\columnwidth]{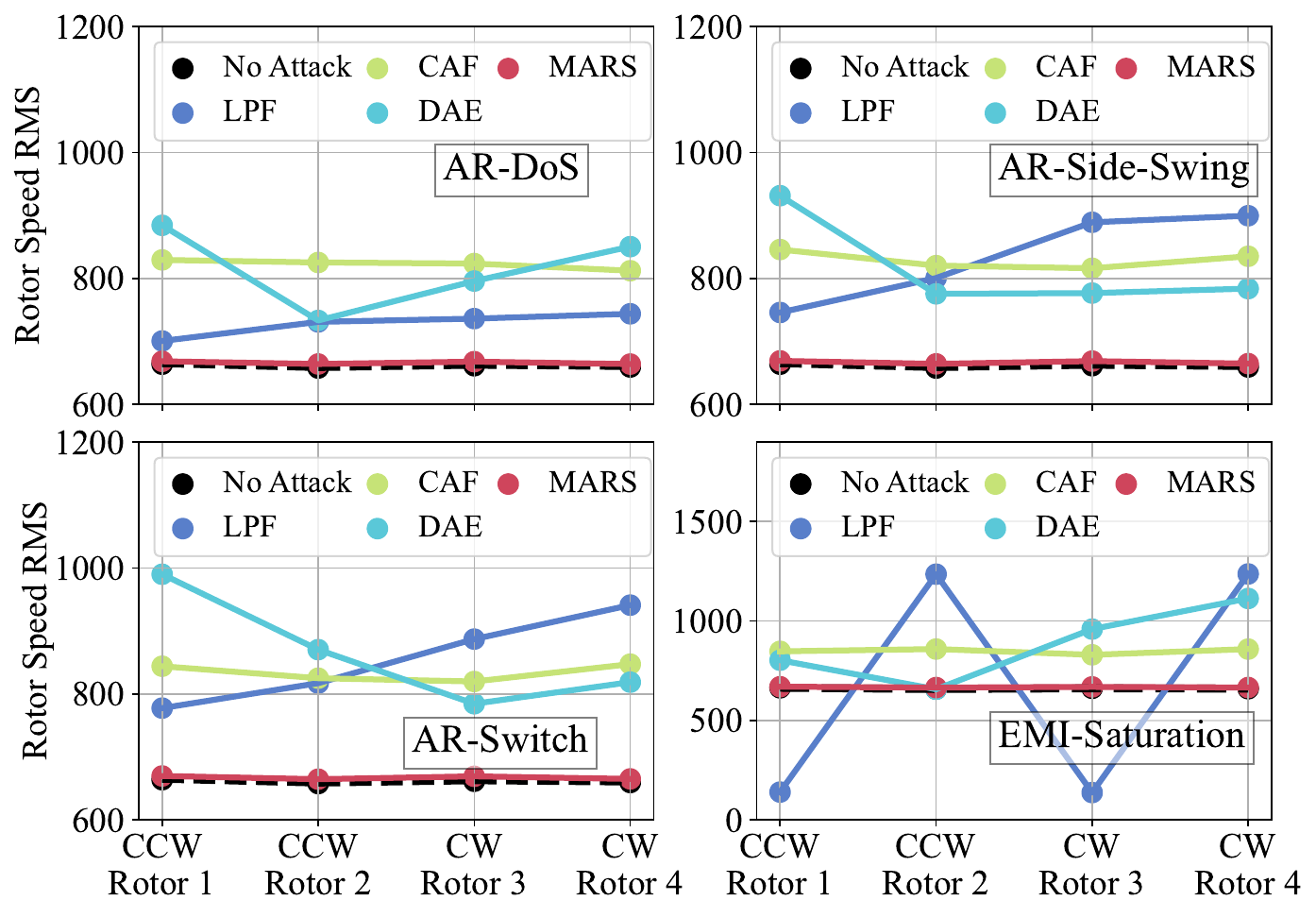}
\caption{Average rotor speed fluctuation while %Root Square Mean (RMS)  in 
hovering.}
\label{fig:hover_rotor_rms}
\end{figure}

For more comprehensive comparisons, we %conducted simulations involving 
considered 4 attacks $\times$ 4 recoveries $\times$ 10 hovering missions, resulting in 160 trials. This generated approximately 160 trials $\times$ $20~s$ per trial $\times$ $250~Hz$ sampling rate = 800k data points. In this evaluation, we focused on the \textit{survival time} and \textit{control smoothness} of the UAV by monitoring the time of staying in air under different attacks and the rotor speed fluctuations from the attack start to the UAV crash time. Table~\ref{tab:hover_survival_time} summarizes the average survival times for the evaluated methods, and Fig.~\ref{fig:hover_rotor_rms} shows the average rotor speed fluctuations. 

\begin{table}[!t]
    % \captionsetup{skip=10pt}
    \caption{Average Survival Time in Hovering Mission.}
    \centering
    \renewcommand{\arraystretch}{1.3}
    \begin{adjustbox}{width=\columnwidth}
    \begin{tabular}{c c c c c}
    % \hhline{=====}
    \hhline{-----}
    \multirow{2}{*}{\makecell[c]{\textbf{Recovery}\\\textbf{Methods}}} & \multicolumn{4}{c}{\textbf{Hovering Mission Average Survival Time ($s$)}} \\ 
    \cline{2-5}
    \cline{2-5}
     & \makecell[c]{AR-DoS} & \makecell[c]{AR-Side-Swing} & \makecell[c]{AR-Switch} & \makecell[c]{EMI-Saturation} \\
    \hhline{-----}
    LPF & 2.826 & 1.463 & 1.252 & 1.283  \\ \hline
    DAE & 2.097 & 2.228 & 1.781 & 1.086  \\ \hline
    CAF & 1.812 & 2.041 & 2.033 & 2.388  \\ \hline
    MARS & Survived & Survived & Survived & Survived  \\ \hline
    % \hhline{=====}
    \hhline{-----}
    \end{tabular}
    \end{adjustbox}
    \label{tab:hover_survival_time}
\end{table}

As shown, \textbf{all} previous methods %struggle to 
were not able to keep the drone in the air %in the presence of attacks on inertial sensing, 
under IMU attacks, with CAF slightly outperforming the other two (detailed discussion is provided in App.~\ref{appendix:hovering_control}). On the other hand, MARS-based control design kept the UAV flying -- 
\emph{MARS is capable of providing timely and reliable control outputs to survive all types of IMU attacks.} MARS successfully survived all four analyzed attacks by immediately isolating the compromised sensors and acquiring body-frame attitude information through a robust sensor fusion that utilizes rotor speed measurements. MARS overcomes the limitations of filtering-based methods, as its performance is much less dependent on the specific profile of attack vectors. As an attitude reconstruction method, it outperforms CAF by leveraging reliable %source of 
rotor speed measurements, %which is 
fused in the EKF to capture subtle changes in drone attitude and provide accurate~control~feedback.

% to the controller.

\subsection{Dynamic Mission Recovery}
\label{subsec:move_analysis}

\begin{figure}[!t]
\centering
\includegraphics[width=\columnwidth]{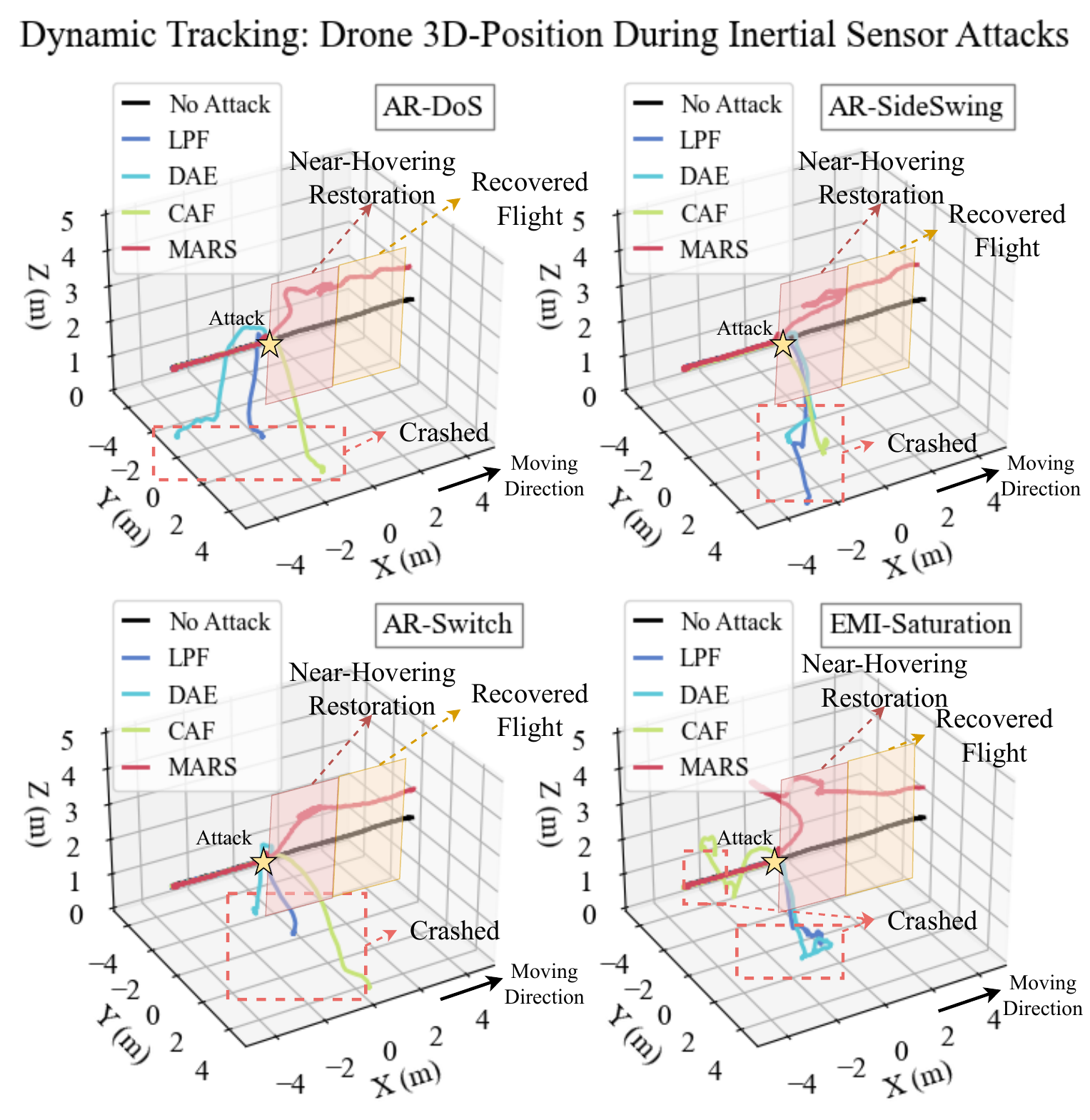}
\caption{3D drone position in a dynamic tracking~mission under 4 IMU sensor attacks, for different recovery~methods.}
\label{fig:moving_3d_plot}
\end{figure}

We further tested MARS multi-stage dynamic recovery strategy in more dynamic flight scenarios (i.e., beyond hovering). %In this analysis, the quadcopter 
The UAV was tasked with completing a tracking mission along a $10~m$ trajectory on the X-axis in the Earth frame while maintaining a fixed altitude. Attacks on the inertial sensors were initiated mid-mission, and our objective was to complete the mission without crashing. Fig.~\ref{fig:moving_3d_plot} illustrates the 3D position of the drone in the dynamic tracking missions. Compared to the three existing methods, which all failed, 
MARS successfully implemented the multi-stage dynamic recovery process, which included `braking' when the attacks were detected, restoring to a near-hovering condition, and then continuing the mission in the recovered flight mode.

\begin{figure}[!t]
\centering
\includegraphics[width=0.976\columnwidth]{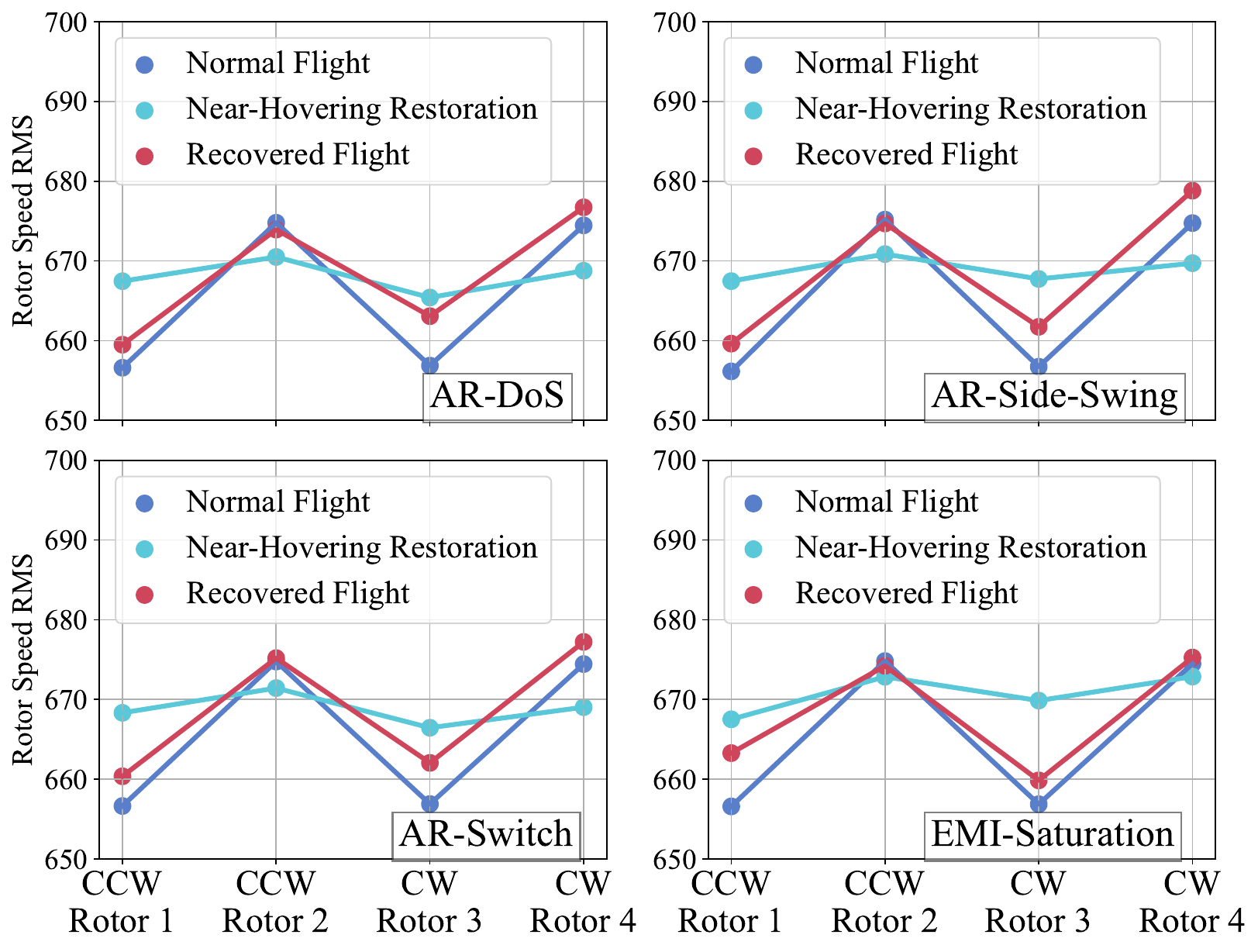}
\caption{Average rotor speed fluctuation in RMS in the multi-stage dynamic recovery of dynamic tracking missions.}
\label{fig:move_rotor_rms}
\end{figure}

\begin{table}[t!]
    % \captionsetup{skip=10pt}
    \caption{Average Tracking Error %(RMSE)
    in Dynamic Missions.}
    \centering
    \renewcommand{\arraystretch}{1.3}
    \begin{adjustbox}{width=\columnwidth}
    \begin{tabular}{c c c c c}
    % \hhline{=====}
    \hhline{-----}
    \multicolumn{2}{c}{} \makecell[c]{\textbf{Dynamic Tracking Mission}} & \multicolumn{3}{c}{\textbf{Lateral Tracking Error ($m$) (RMSE)}} \\
    \hline
     Attack Vector & Recovery Method  & {No Attack} & {Recovered} & {Increment}\\
    \hhline{-----}
    AR-DoS & MARS & 0.256 & 0.459 & 0.203 \\ \hline
    AR-Side-Swing & MARS & 0.256 & 0.454 & 0.198 \\ \hline
    AR-Switch & MARS & 0.256 & 0.362  & 0.106 \\ \hline
    EMI-Saturation & MARS & 0.256 & 0.638 & 0.382 \\
    % \hhline{=====}
    \hhline{-----}
    \end{tabular}
    \end{adjustbox}
    \label{tab:move_rotor_rmse}
\end{table}

\begin{table}[t!]
    % \captionsetup{skip=10pt}
    \caption{Average Completion Time in Dynamic Tracking.} % Mission.}
    \centering
    \renewcommand{\arraystretch}{1.3}
    \begin{adjustbox}{width=\columnwidth}
    \begin{tabular}{c c c c c}
    % \hhline{=====}
    \hhline{-----}
    \multicolumn{5}{c}{\textbf{Dynamic Tracking Mission Completion Time ($s$)}} \\
    \hline
     No Attack & AR-DoS & AR-Side-Swing & AR-Switch & EMI-Saturation \\ \hline
     22.772 & 28.212 & 27.968 & 28.113 & 28.769 \\ \hline
    % \hhline{=====}
    \hhline{-----}
    \end{tabular}
    \end{adjustbox}
    \label{tab:move_mission_time}
\end{table}

To evaluate MARS multi-stage dynamic recovery, we analyze the recovered flights from the perspective of the \textit{control smoothness} of the UAV movement reflected by the rotor speed fluctuations, and whether the \textit{mission was completed}, % of the tracking task, 
reflected by (a) the position tracking error in the lateral direction in RMSE, and (b) mission completion time.
Our results indicate that MARS maintains rotor speed fluctuations at a low level, with the near-hovering restoration stage experiencing the smallest oscillation. %, effectively mitigating the impact of the attack. 
The control smoothness of the normal flight was largely preserved in the recovered flight, with a rotor speed RMS difference of less than $10~rad/s$. From the mission completion perspective, MARS achieved an average tracking error of approximately 0.5 $m$ in the lateral direction, and the increase in completion time was below 30\%, as shown in Tables~\ref{tab:move_rotor_rmse} and~\ref{tab:move_mission_time}.

Thus, 
\emph{the MARS multi-stage dynamic recovery %maintains control effectiveness
provides attack-resilience, both in terms of the control quality as well as mission completion time.} Although the braking mechanism temporarily suspends the drone's flight, the multi-stage dynamic recovery prevents significant control oscillations or UAV crashes. Also, the drone maintains control efficiency, completing the task with an acceptable tracking error.

%% file: Experiment.tex
\section{Real-world Case Studies}
\label{sec:physical}

% To test MARS in real-world scenarios, we 
We built a physical platform using commercially available quadrotor frames and onboard sensors, and deployed the modified MARS-PX4 autopilot. Demo videos are available at~\cite{MARS}.

% \subsection{Physical Platform Design}

\begin{figure}[!t]
    \centering
    \includegraphics[width=0.928\columnwidth]{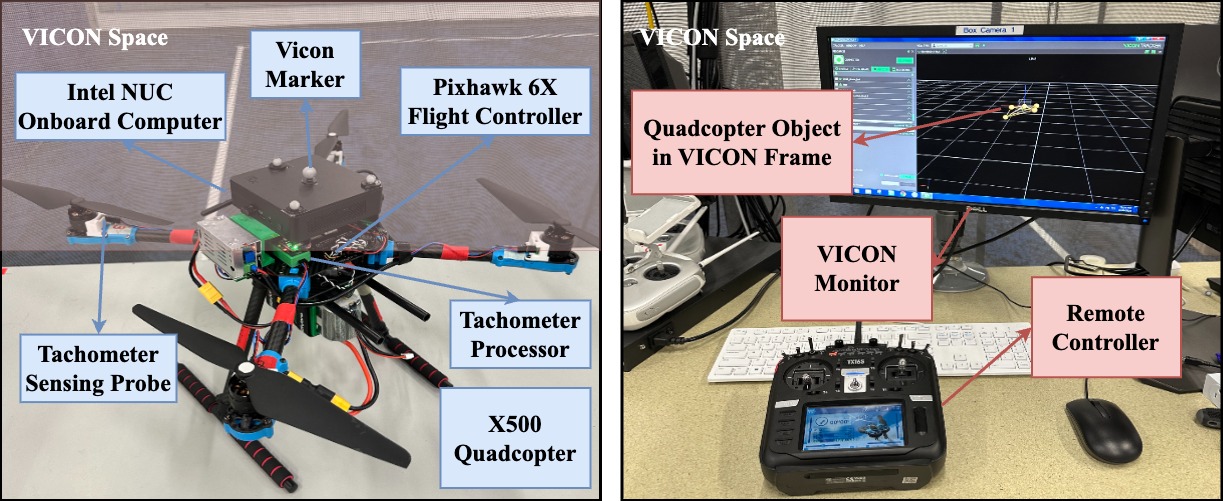}
    \caption{MARS-PX4 X500 physical platform setup.}
    \label{pic:x500}
\end{figure}

\vspace{2pt}
\noindent
\textbf{Experimental Setup (Fig.~\ref{pic:x500}).}
We built a customized quadrotor with Pixhawk 6X flight controller and Holybro X500 V2 airframe, as well as an Intel NUC computer to generate high-level commands and record datasets. For safety concerns, we conducted experiments first in an indoor arena with VICON motion capture system (MOCAP) installed. % which was used as a graound truth.

Initial onboard sensors included an IMU, a compass, and a barometer. To enable MARS, we added four ThunderFly TFRPM01 I2C bus-based drone tachometer sensors to the Pixhawk 6X board. Moreover, VICON MOCAP played two roles in our experiments: it was used to simulate GPS measurements and compass for indoor environments at a frequency of $10~Hz$, and to provide ground-truth data for analysis at $200~Hz$. 
For MARS-based resilient control, \textit{only MOCAP (to emulate GPS) % and compass), 
and tachometer} measurements were utilized to generate state estimates for the drone controllers.
The details on the sensing parameters (e.g., update frequency and data output), as well as the parameters of the autopilot modules are provided in App.~\ref{appendix:system_params} (Tables~\ref{tab:x500_sensors},~\ref{tab:x500_controllers}).

\vspace{3.2pt}
\noindent\textbf{MARS-PX4 Computation Overhead.}
We measured the computation overhead of MARS modules inside the PX4 ecosystem. On average, MARS only takes $0.261~ms$ to execute with minimal increase in CPU ($1\%$) and memory usage ($12~KB) $ %. More instrumentation measurements are provided 
(more results in App.~\ref{app:overhead} (Table~\ref{tab:mars_cpu_and_memory_usage} and Fig.~\ref{fig:mars_execution_time}). %-- showing that MARS finishes online execution within $0.3~ms$.

% \subsection{Real-time Resilient State Estimation}

\subsection{Real-time Attacks and Anomaly Detection}

% \paragraph{Tachometer-based estimation of control inputs.}
\noindent\textbf{Tachometer-based estimation of thrust and torque.}
In App.~\ref{app:tachometer_physical}, our experimental results show that with tachometer data sampled at $200~Hz$ on X500, we accurately estimate {thrust and torque} from high quality rotor speed measurement, enabling resilient state estimate without~IMUs.

\begin{figure}[t!]
    \centering
    \subfloat%[Euler angle estimate error.]
    {\includegraphics[width=0.50\columnwidth]{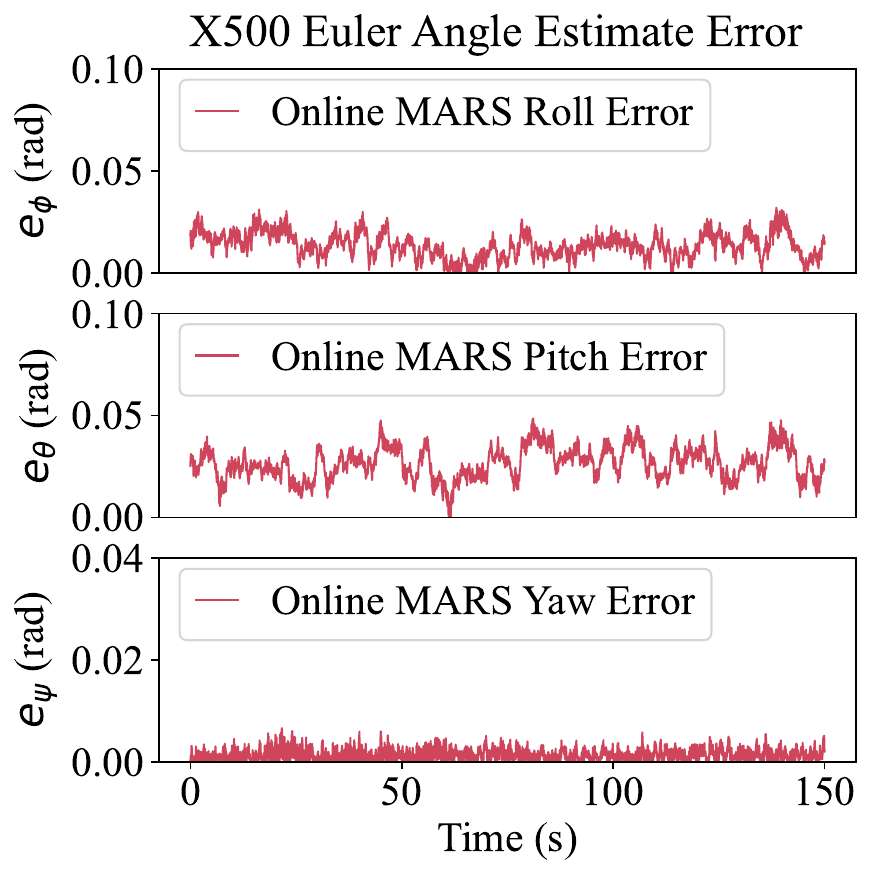}}
    \hfil
    \subfloat%[Ang. velocity estimate error.]
    {\includegraphics[width=0.50\columnwidth]{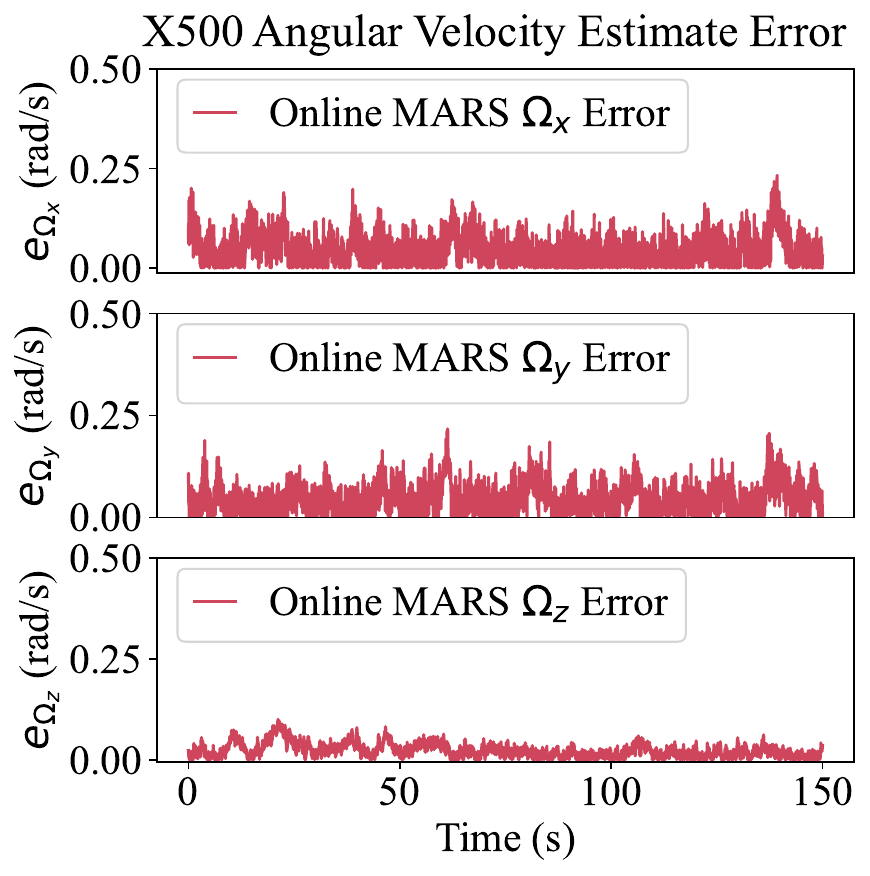}}
    \caption{Online attitude and angular velocity estimate error of MARS running on the X500 quadrotor.}
    \label{fig:x500_estimate_error}
\end{figure}

\vspace{3.2pt}
\noindent\textbf{MARS real-time resilient state estimate.}
We also evaluated the estimation accuracy of online MARS before feeding the estimates into the controller. We set the drone into hovering and focused on MARS-RSE error from the %MOCAP data which served as 
ground truth; %(see %As shown in 
from Fig.~\ref{fig:x500_estimate_error},  % shows the estimate error in Euler angles and angular velocities produced by online MARS resilient state estimator.
%
%The plots indicate that 
the error level of online %state estimates 
MARS-RSE is comparable to the one in SITL environments from Fig.~\ref{fig:angular_estimate_error}. The results highlight the consistent MARS-RSE performance both in simulations and the real-world experiments.

% \begin{table}[t!]
%     % \captionsetup{skip=10pt}
%     \caption{Computation overhead of MARS-PX4.}
%     \centering
%     \renewcommand{\arraystretch}{1.3}
%     \begin{adjustbox}{width=\columnwidth}
%     \begin{tabular}{c c c c}
%     % \hhline{====}
%     \hhline{----}
%     \multicolumn{4}{c}{\textbf{MARS-PX4 CPU Work Queue (WQ) Usage (\%)}} \\
%     \hline
%      CPU Usage & WQ without MARS & WQ with MARS & Increment \\ \hline
%      $\sim$47 & 0.188 & 1.499 & 1.311 \\ \hline
%     \hhline{----}
%     % \hhline{====}
%     \multicolumn{4}{c}{\textbf{MARS-PX4 Memory Usage ($KB$)}} \\
%     \hline
%      Total Memory & Usage without MARS & Usage with MARS & Increment \\ \hline
%      916.416 & 365.840 & 377.808 & 11.968 \\ \hline
%     \hhline{----}
%     % \hhline{====}
%     \multicolumn{4}{c}{\textbf{MARS-PX4 Average Execution Time ($ms$)}} \\
%     \hline
%      Total & Initialization & MARS-RSE & MARS-AD \\ \hline
%      0.261 & 0.029 & 0.200 & 0.032 \\ \hline
%     % \hhline{====}
%     \hhline{----}
%     \end{tabular}
%     \end{adjustbox}
%     \label{tab:mars_cpu_and_memory_usage}
% \end{table}

% \subsection{Real-time Attacks and Anomaly Detection}

\begin{figure}[t!]
    \centering
    \subfloat%[MARS real-time angular velocity recovery.]
    {\includegraphics[width=0.514\columnwidth]{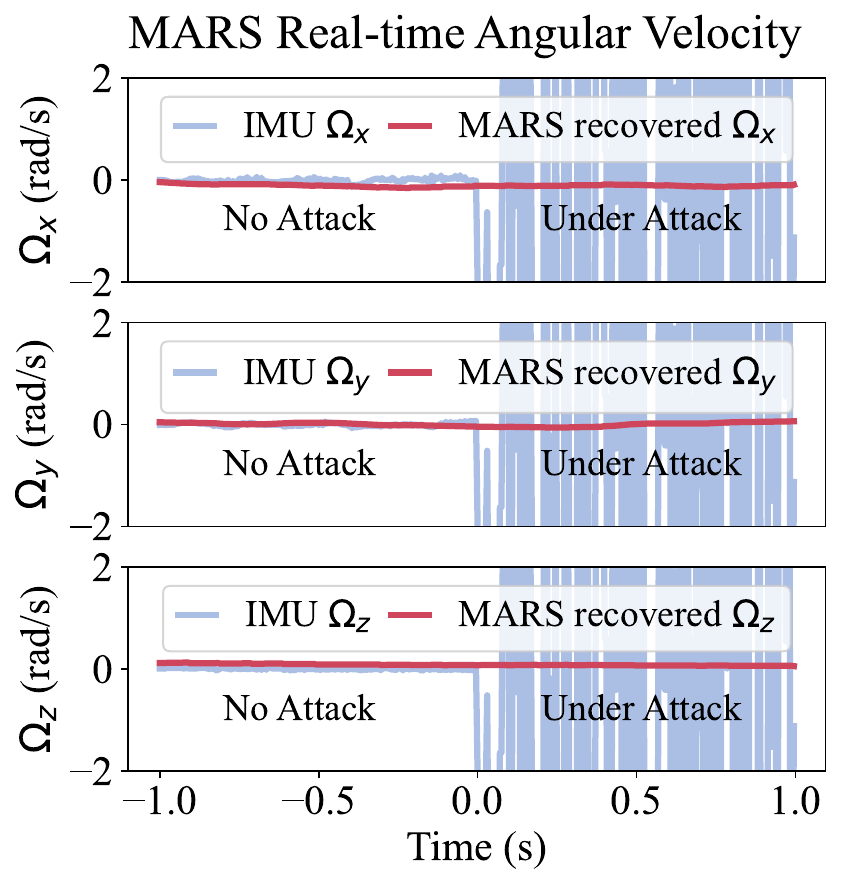}}
    \hfil
    \subfloat%[MARS real-time anomaly detection.]
    {\includegraphics[width=0.486\columnwidth]{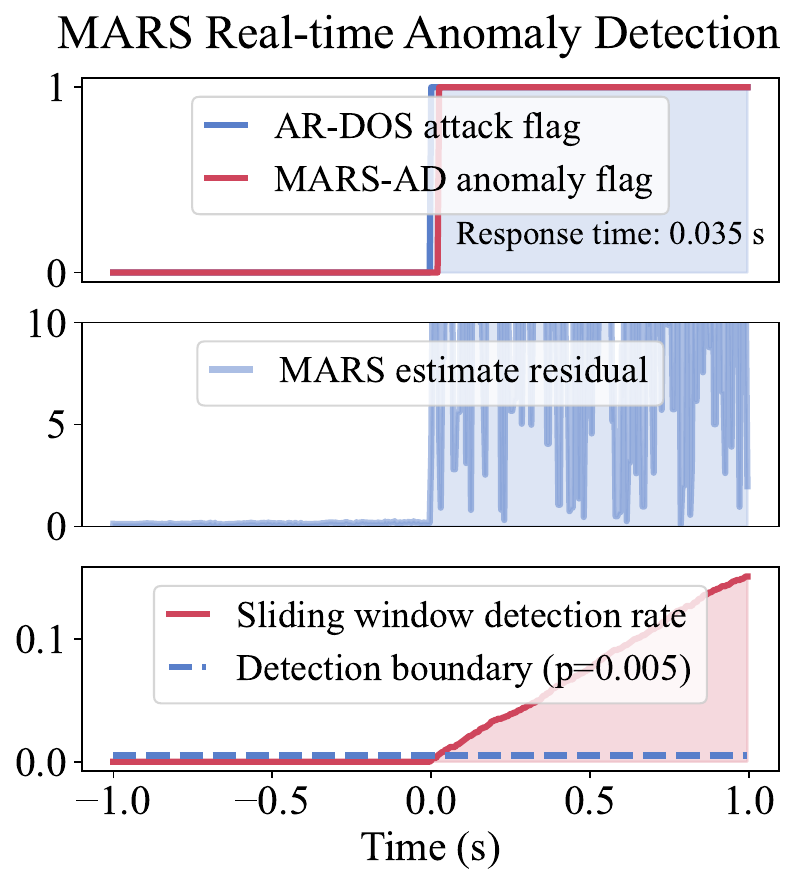}}
    \caption{Real-time MARS signal recovery and anomaly detection results from the X500 quadrotor.}
    \label{fig:online_recovery_and_ad}
\end{figure}

\vspace{3.2pt}
\noindent\textbf{Online attack generation.} {In our experiments, we used realistic attack vectors %with AR-DoS attack modulation as 
discussed in Section~\ref{subsec:ad_analysis} into PX4 firmware. % to evaluate the efficacy of \emph{MARS anomaly detection}, and \emph{MARS resilient control performance}.} %of the drone using MARS state estimates.

\vspace{3.2pt}
\noindent\textbf{Online detection results.}
Fig.~\ref{fig:online_recovery_and_ad} %depicts the process of 
shows %online 
attack detection %at the time of attack 
when the attack starts at $t_{attack}=0$; %where 
MARS-AD identifies the %anomalous behavior in IMU sensor measurements 
IMU attacks with a significant increase in MARS estimate residual and sliding window detection rate. The %system anomaly is declared  
attack is detected $0.035~s$ after it started.
Table~\ref{tab:real_time_anomaly_detection_times} shows the average detection time for 4 different attack profiles (10 attacks each) in real flights. The statistical results demonstrate  %the adaptability of 
MARS-AD effectiveness in both simulated and real-world scenarios.

\begin{table}[!t]
    % \captionsetup{skip=10pt}
    \caption{MARS-AD average response time in real flights.}
    \centering
    \renewcommand{\arraystretch}{1.3}
    \begin{adjustbox}{width=\columnwidth}
    \begin{tabular}{c c c c}
    % \hhline{====}
    \hhline{----}
     \multicolumn{4}{c}{\textbf{MARS Anomaly Detector Average Response Time ($s$)}} \\ 
    \cline{1-4}
    \cline{1-4}
     \makecell[c]{AR-DoS} & \makecell[c]{AR-Side-Swing} & \makecell[c]{AR-Switch} & \makecell[c]{EMI-Saturation} \\
    \hhline{----}
    0.030 & 0.065 & 0.026 & 0.024  \\ \hline
    % \hhline{====}
    \hhline{----}
    \end{tabular}
    \end{adjustbox}\label{tab:real_time_anomaly_detection_times}
\end{table}

\subsection{Recovering Real-world Flights}
\label{subsec:mars_real_world_flights}

% \noindent\textbf{MARS real-world flight design.} 
% We tested the full MARS %detection and recovery 
% framework on the MARS-PX4 X500 quadroptor in real flights. 
We conducted four \emph{real-world} flight case studies:
$(i)$~Hovering; $(ii)$~Way-point visiting; $(iii)$~Square tracking; $(iv)$~Hovering under wind disturbance. In %the hovering experiment
$(i)$ experiments, %MARS's ability to detect consecutive attacks and control the drone without crashing was tested against simulated EMI attack vectors. 
a series of impulsive EMI saturation attacks of $0.5~s$ per event were created~\cite{8986669} to introduce sudden perturbations to test the responsiveness of MARS-AD and the smoothness of the controller \emph{switching between standard and resilient modes}.
%but is also a realistic setting for attackers employing directional spoofing devices on agile UAVs}.  
In %way-point visiting and square tracking
$(ii)$ and $(iii)$, we demonstrated the dynamic recovery capability in surviving attacks and completing missions against AR-DoS~attacks. To further demonstrate MARS's robustness to dynamical changing environments, we created \emph{wind disturbance} using a wind blower testing MARS against attacks under this challenging condition. The wind blower was set directly pointing at the drone from a distance of $3~m$, % at the same height of UAV flying, 
and produced a wind disturbance at around $3~m/s$ at the hovering position (measured by a handheld anemometer).

% \begin{figure*}[t!]
%     \centering
% {\includegraphics[width=0.82\textwidth]{Figures/Experiment Figures/mars_flight_tests_1114.drawio.pdf}}
%     \caption{MARS real flight tests on X500 quadcopter. Left: Hovering; middle: Way-point visiting; right: Square tracking.}
%     \label{fig:x500_real_flights}
% \end{figure*}

\begin{figure*}[t!]
    \centering
{\includegraphics[width=\textwidth]{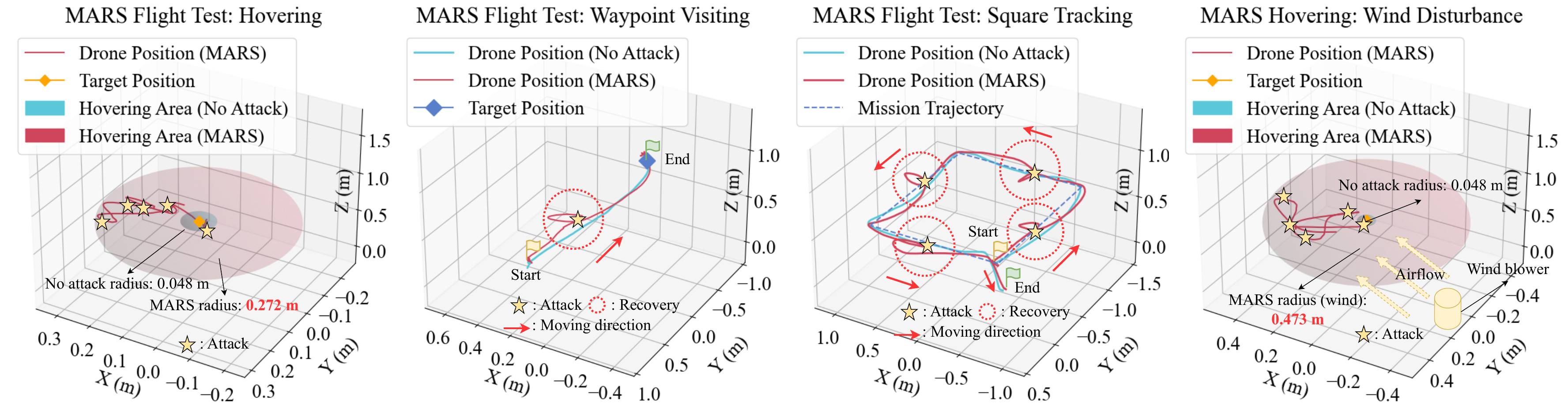}}
    \caption{MARS real flight tests. Left to right: Hovering; Way-point visiting; Square tracking; Hovering under disturbance.}
    \label{fig:x500_real_flights}
\end{figure*}

\vspace{3.2pt}
\noindent\textbf{MARS successfully defended against attacks.} Fig.~\ref{fig:x500_real_flights} captures the drone trajectories, %of the drone in the case studies, indicating 
showing successful attack detection and recovery in various missions. In the hovering case, MARS detected the %destructive 
EMI saturation attack signals and switched to resilient control, making the drone hover near the target position, with the largest deviation at $0.272~m$. In the dynamic missions, the multi-stage recovery strategy was effectively executed, pausing the movement of drone until attack is cleared, before continuing with the mission.

% \begin{table}[!t]
%     \caption{MARS rotor speed fluctuations in standard deviation.}
%     \centering
%     \renewcommand{\arraystretch}{1.3}
%     \begin{adjustbox}{width=\columnwidth}
%     \begin{tabular}{c c c c c}
%     \hhline{=====}
%      \multicolumn{5}{c}{\textbf{MARS Real-flight Rotor Speed Fluctuations (rad/s)}} \\ 
%     \cline{1-5}
%     \cline{1-5}
%      \makecell[c]{Status} & \makecell[c]{$\sigma_{\omega_1}$ (CCW)} & \makecell[c]{$\sigma_{\omega_2}$ (CCW)} & \makecell[c]{$\sigma_{\omega_3}$ (CW)} & \makecell[c]{$\sigma_{\omega_4}$ (CW)}\\
%     \hhline{=====}
%     No Attack & 19.247 & 17.134 & 20.083 & 16.573 \\ \hline
%     Recovered & 24.960 & 15.379 & 24.958 & 21.611 \\ \hline
%     \hhline{=====}
%     \end{tabular}
%     \end{adjustbox}\label{tab:mars_flight_rotor_speeds}
% \end{table}

\begin{table}[!t]
    \caption{MARS real-flight Euler angle fluctuations ($rad$).}
    \centering
    \renewcommand{\arraystretch}{1.3}
    \begin{adjustbox}{width=\columnwidth}
    \begin{tabular}{c c c c c c}
    % \hhline{======}
    \hhline{------}
    % \multicolumn{6}{c}{\textbf{MARS Real-flight Euler Angle Fluctuations ($rad$)}} \\
    % \hhline{------}
    \multirow{2}{*}{Mission} & 
    \multirow{2}{*}{Status} & 
    \multicolumn{2}{c}{Roll Angle} & 
    \multicolumn{2}{c}{Pitch Angle} \\
    \hhline{~~----}
     & & 
    Max Tilt & 
    Std. & 
    Max Tilt & 
    Std. \\
    \hhline{------}
    \multirow{2}{*}{Hovering} & 
    No Attack & 0.012 & 0.003 & 0.063 & 0.002 \\
    \hhline{~-----}
     & 
    MARS & 0.053 & 0.017 & 0.297 & 0.068  \\
    \hhline{------}
    \multirow{2}{*}{\makecell[c]{Way-point\\Visiting}} & 
    No Attack & 0.043 & 0.007 & 0.031 & 0.003 \\
    \hhline{~-----}
     & 
    MARS & 0.184 & 0.015 & 0.152 & 0.013 \\
    \hhline{------}
    \multirow{2}{*}{\makecell[c]{Square\\Tracking}} & 
    No Attack & 0.036 &  0.007 & 0.077 & 0.013 \\
    \hhline{~-----}
     & 
    MARS & 0.190 & 0.019 & 0.405 & 0.030 \\
    \hhline{------}
    \makecell[c]{Hovering (Wind)}& 
    MARS & 0.234 & 0.058 & 0.248 & 0.073 \\
    % \hhline{======}
    \hhline{------}
    \end{tabular}
    \end{adjustbox}
    \label{tab:euler_angle_fluctuations}
\end{table}

\vspace{3.2pt}
\noindent
\textbf{MARS minimized control performance loss through dynamic recovery.} Table~\ref{tab:euler_angle_fluctuations} shows the maximum tilt and the standard deviation of roll and pitch angles measured by the motion capture system. %during the missions. 
Due to the near-hovering assumption, MARS had good control performance in hovering, with the maximum tilt in pitch angle at $0.297~rad$. The dynamic missions experienced acute behavior when MARS executed emergency braking, with the maximum tilt $0.184~rad$ in roll angle in way-point visiting mission, and $0.405~rad$ in pitch angle in square tracking case. %{The reason for the larger maximum tilt angle observed in the hovering case than the way-point visiting, is that in the former we applied a longer duration and consecutive attacks, compared to the attack used in the latter scenario}. 
Nevertheless, MARS maintained a low oscillation, showing stable mission execution. % in the long run. 

\noindent\textbf{MARS demonstrated robustness to dynamical environments.} Fig.~\ref{fig:x500_real_flights} and Table~\ref{tab:euler_angle_fluctuations} show that despite the increment of position deviation to $0.473~m$, with maximum tilts in roll and pitch angles $0.234~rad$ and $0.248~rad$ due to the wind, MARS managed to control the drone with acceptable stability. 
We also conducted \emph{outdoor} tests; %using an actual GPS} 
videos provided at~\cite{MARS}.

%% file: Discussion.tex
\section{Discussion and Conclusion}
\label{sec:future}

This paper  introduced MARS, a model-based anomaly detection and recovery system that secures UAVs from attacks on inertial sensors. By isolating compromised IMUs and estimating {thrust and torque} from rotor speed measurements, a resilient state estimator reconstructs the missing body-frame attitude and angular velocity %information 
for the drone controller. We demonstrated the capability of MARS to achieve stable hovering and dynamical tracking by accurately detecting realistic attack vectors and actively executing the multi-stage dynamical flight recovery strategy. MARS is designed to be %a general anomaly detection and recovery framework,  
universally applicable for UAVs encountering different threats %models 
on inertial sensors. We now discuss %the assumptions, limitations and future scopes. % of MARS.
MARS limitations and~future~work. 

% \paragraph{Near-Hovering Assumption for Control Input Estimate.}
\vspace{2pt}
\noindent\textbf{The Near-Hovering Assumption.} % for Control Input Estimate}
% Having access to accurate control inputs is a critical part for Kalman filter-based state estimators. With trustworthy inertial sensors, the control input to drone system is the body-frame acceleration and angular velocity measurements. When they become unavailable due to various attacks, we adopt tachometers to acquire propeller revolution speed and estimate the direct control input to the drone system, which is the collective torque and thrust applied at center of gravity. Mathematically, this estimation also needs body-frame angular velocity information as is shown in Eq.~\eqref{eq_full_control_input}. Eqs.~\eqref{eq_full_control_input}-\eqref{eq_simplified_control_input} achieve an approximation that ignores the effect of angular velocity and produces an estimate using only rotor speeds and earth frame velocity measurements.
%
% However, this 
It causes control estimate degradation when the drone experiences acute behaviors, and may lead to an increase in the estimation error. % in state estimators. 
% This scenario typically happens in the small time window, from the attack start %ing from the time of attack 
% until the detection and control mode switch. Within this small time-period, % of time, 
% a powerful attack %vector 
% could cause a deviation in the attitude of the drone, moving it further away from the near-hovering point. 
We addressed this problem by: %three parts:
$(i)$~a fast AD reducing the deviation to the minimum;
$(ii)$~a temporary `braking' mechanism pausing the movement and restoring the UAV to near-hovering;
$(iii)$~an appropriate torque compensation to facilitate the braking process. MARS aims to minimize the impact caused by inaccurate state estimates due to the deviation from near-hovering conditions under different dynamical missions. While it secures a moving drone from crashing, it sacrifices the control performance, %leaving it under a 
with resilient but restricted operation. This is an intrinsic limitation of rotor speeds-based {thrust and torque} estimation when inertial sensors cannot be used.

\vspace{2pt}
\noindent\textbf{Adaptability to Dynamical Environments.}
We demonstrated MARS robustness to wind disturbances. Such dynamic environmental changes may lead to increased estimation errors, % in body-frame information, 
thereby degrading control performance. To enhance MARS's adaptability to changing environmental conditions, also integrating  %a mathematical model 
an unknown input observer 
for disturbances would be beneficial; e.g., the Dryden Model~\cite{DoD2012FlyingQualities}, as in~\cite{quinonez2020savior}, enables accurate modeling of drone dynamics under wind influence. Also, learning-based system identification methods~\cite{1643442, yu2017preparing} could be used to adaptively update MARS system parameters, improving the accuracy of the resilient~estimation.

% \vspace{3.2pt}
% \noindent\textbf%{Towards Identifying and 
% {Recovering from Multi-Modality Sensor Attacks.}
% While MARS %manages to detect and recover the attack vectors on 
% provides resiliency against IMU attacks, %on inertial sensors, %enormous work have demonstrated 
% in future we plan to address the reported vulnerabilities of other onboard sensors  (e.g., GPS, radar~\cite{hunt_ndss24} or Lidar~\cite{hallyburton_security22} spoofing). 
% Attacks on multi-modality sensing create a coupling effect with more complicated patterns. Thus, the correct identification of the attack source and victim sensor %establishes a solid basis for further 
% is critical for recovery. To achieve this, a thorough analysis of attack signatures for perception-based attack vectors needs to be done, to enable ADs to identify the unique attack characteristics. % of a certain attack profile. 

%% file: Appendix.tex
\section{Appendix}
\label{sec:appendix}

% \subsection{Nonlinear Vehicle Model}
\input{System_Model}

\subsection{Attacks on Inertial Sensors}

% This section discusses the details of acoustic resonant attack and electromagnetic interference attack.

\subsubsection{Acoustic Resonant Attack}
\label{subsec:appendix_acoustic}

% The 
We summarize the main properties of acoustic resonant attacks primarily include mathematical formulation, sampling drift effect, attacker manipulation, and attack requirements and effects~\cite{tu2018injected}.

\noindent\textbf{Attack formulation}: When the sensing mass is oscillating under the effect of acoustic injection, the digitized (i.e., sampled) sensor output signal  $V[i]$ is represented by:
\begin{equation}
\begin{split}
V[i] = A \cdot \sin(2 \pi \nu \frac{i}{F_S} + \phi_0),\quad
\nu = F - n \cdot F_S,
\end{split}
\end{equation}
where $A$ and $\phi_0$ are the amplitude and phase of the oscillating signal, respectively; $\nu$ refers to the aliased frequency due to the Nyquist sampling theorem, where $F$ is the frequency of the analog signal triggering the acoustic resonance, $F_S$ is the sampling frequency, and $n$ is the integer that describes the aliasing relationship between them.

% %
% \begin{equation}
% \begin{split}
% \hat{a}_{acc}[i] &= {a}_{acc}[i] + A_{acc} \cdot \sin\left(2\pi \epsilon_{acc} \frac{i}{F_S} + \phi_{0,acc}\right) \\
% \hat{\Omega}_{gyro}[i] &= {\Omega}_{gyro}[i] + A_{gyro} \cdot \sin\left(2\pi \epsilon_{gyro} \frac{i}{F_S} + \phi_{0,gyro}\right)
% \end{split}
% \end{equation}
% %

\noindent\textbf{Sampling drift}: While the mathematical formulation of the acoustic attack is well-patterned, simply cutting off at the frequency $\frac{\nu}{F_S}$ is not sufficient to filter out the attack vectors. The reason is that the sampling interval of a real-time system is not perfectly maintained at $\frac{1}{F_S}$. A slight drift in the sampling rate, $\Delta F_S$, can be amplified by $n$ times into $\Delta \nu$ due to the aliasing effect:
\begin{equation*}
\begin{split}
\Delta \nu = \left(F - n \cdot (F_S + \Delta F_S)\right) - \left(F - n \cdot F_S\right)
= - n \cdot \Delta F_S.
\end{split}
\label{eq_sampling_drift}
\end{equation*}
The typical resonant frequency of IMU sensors ranges from $20~kHz$ to $30~kHz$, while the sampling frequency is often around $1~kHz$ %. Eq~\eqref{eq_sampling_drift} 
-- the above equation indicates that a shift of $0.01~Hz$ in the sampling rate will be amplified $100$ times into $1~Hz$ in the induced frequency.

\noindent\textbf{Attacker manipulation}: The attacker is also capable of manipulating the amplitude and frequency of the acoustic resonant signals. The compromised sensor measurement is a combination of the benign signal and the acoustic resonance
%\
\begin{equation*}
\begin{split}
\tilde{s}[i] = {s}[i] + A[i] \cdot \sin(\Phi[i]), \quad
\Phi[i] = 2 \pi \nu[i] \frac{i}{F_S} + \phi_0,
\end{split}
\end{equation*}
where $\tilde{s}[i]$ and ${s}[i]$ are the compromised and benign sensor readings from either the accelerometer or gyroscope due to the effects of acoustic resonant attacks. $A[i]$ and $\Phi[i]$ are the modulated amplitude and phase at the aliased frequency $\nu[i]$. The attacker takes advantage of amplitude tuning and phase shifting to induce different acoustic attack variations. The commonly used variations include Denial-of-Service (DoS), Side-Swing, and Switch attacks. These variations pose significant challenges for defense mechanisms in detection, identification, and mitigation.

% %
% \begin{equation}
% \begin{split}
% V[i] = A[i] \cdot \sin(\Phi[i]), \quad
% \Phi[i] = 2 \pi \epsilon[i] \frac{i}{F_S} + \phi_0
% \end{split}
% \end{equation}
% %

% %
% \begin{equation}
% \begin{split}
% \hat{a}_{acc}[i] &= {a}_{acc}[i] + A_{acc}[i] \cdot \sin\left(2\pi \epsilon_{acc}[i] \frac{i}{F_S} + \phi_{0,acc}\right) \\
% \hat{\Omega}_{gyro}[i] &= {\Omega}_{gyro}[i] + A_{gyro}[i] \cdot \sin\left(2\pi \epsilon_{gyro}[i] \frac{i}{F_S} + \phi_{0,gyro}\right)
% \end{split}
% \end{equation}
% %

%  where ${a}_{acc}[i]$ and $\Omega_{gyro}$ are the benign sensor readings from IMU acclerometer and IMU gyroscope. $\hat{a}_{acc}[i]$ and $\hat{\Omega}_{gyro}[i]$ are the compromised measurements due to the acoustic resonance attack.

\noindent\textbf{Attack requirement and effect}: To successfully carry out an acoustic resonant attack, the attacker first identifies the target inertial sensors to determine their resonant frequencies. Then, using consumer-grade speakers or long-range acoustic devices, the attacker broadcasts an attack signal with proper modulation to strike the MEMS components of the victim sensors. Due to the strong oscillations in inertial sensor measurements caused by the resonance effect, the drone loses reliable attitude estimates within tens of milliseconds and faces the danger of losing stable control.

\subsubsection{Electromagnetic Interference Attack}
\label{subsec:appendix_emi}
% 
% % \textbf{Attack requirement and effect}: 
% In order to 
To successfully launch an EMI attack against a target drone~\cite{jang2023paralyzing}, the attacker needs to identify the unique frequency range of the specific control board that is susceptible to electromagnetic interference and the required signal power to trigger the attack. Then, using antennas and RF generators, a physical attack path can be created to target the victim drone. The attacked communication channel could potentially cause unintended bit flips, incorrect values, sensor saturation, and temporary or persistent data loss. The corrupted data received at the control unit leads to serious oscillations in the sensor fusion module, further causing significant fluctuations in control output and rotor~speeds~\cite{jang2023paralyzing}.

% \textbf{Comparison with acoustic resonant attack}: 
Compared with acoustic resonant attacks, %electromagnetic interference (EMI) 
EMI attacks do not target the sensing properties of the %inertial sensors
IMUs but rather the communication link between them and the flight control unit. Unlike acoustic resonance, which can be mathematically formulated into unique patterns, EMI induces arbitrary %attack 
effects where benign signals are largely overwhelmed or completely blocked due to data packet mismatches, the transmission of incorrect values, and communication losses. 

\subsection{Nonlinear Extended Kalman Filter (EKF)}
\label{appendix:ekf}

This section summarizes the  implementation of an EKF for the nonlinear quadrotor system. Starting from the continuous-time model summarized in Sec.~\ref{appendix:drone_model}, the state-space form of the quadrotor discrete-time dynamics is: %can be presented by:
\begin{equation}\label{eq:sys_cont}
\begin{split}
\mathbf{x}_{k} &= f(\mathbf{x}_{k-1},\mathbf{u}_{k-1})+\mathbf{w}_{k-1},\\
\mathbf{y}_k&= h(\mathbf{x}_k)+\mathbf{v}_k,
\end{split}
\end{equation}
where $\mathbf{w}_{k}$ is the system disturbance, $\mathbf{y}_{k}$ is the vector of the raw sensor measurements with Gaussian noise $\mathbf{v}_{k}$. 

EKF has a two-step process, predict step and update step. In the predict step, the state transition function $f$ can be used to calculate the predicted state $\mathbf{\hat{x}}_{k|k-1}$ from the estimation of previous time $\mathbf{\hat{x}}_{k-1|k-1}$ and control input $\mathbf{u_k}$, and observation function $h$ can be used to calculate the predicted covariance $\mathbf{P}_{k|k-1}$: 
\begin{equation}\label{eq:ekf_pred}
\begin{split}
\mathbf{\hat{x}}_{k|k-1} &= f(\mathbf{\hat{x}}_{k-1|k-1},\mathbf{u}_k),\\
\mathbf{P}_{k|k-1}&= h(\mathbf{F}_{k}\mathbf{P}_{k-1|k-1}\mathbf{F}_k^T + \mathbf{Q}_k);
\end{split}
\end{equation}
here $\mathbf{F}_k$ is the Jacobian matrix of state transition function, $\mathbf{u_k}$ is control input, $\mathbf{P}_{k-1|k-1}$ is previous covariance estimate, $\mathbf{Q}_k$ is the covariance of process noise. The state transition and observation matrices $\mathbf{F_{k}}$ and $\mathbf{H_{k}}$ are %defined as 
the Jacobians of the state transition and observation functions:
\begin{equation}\label{eq:ekf_jacobian}
%\begin{split}
\mathbf{F}_k = 
{\frac{\partial f}{\partial x}|_{\mathbf{\hat{x}}_{k-1|k-1}, \mathbf{u}_{k}}},\qquad
\mathbf{H}_{k} = 
\frac{\partial h}{\partial x}|_{\mathbf{\hat{x}}_{k|k-1}}.
%\end{split}
\end{equation}

In update step, the innovation residual $\mathbf{\Tilde{s}}_k$, innovation covariance $\mathbf{S}_k$, Kalman gain $\mathbf{K}_k$, updated state estimate $\mathbf{\hat{x}_{k|k}}$ and updated covariance estimate $\mathbf{P}_{k|k}$ can be computed:
\begin{equation}\label{eq:ekf_update}
\begin{split}
\mathbf{\Tilde{s}}_k &= \mathbf{y}_k - h(\mathbf{\hat{x}}_{k|k-1}),\\
\mathbf{S}_k &= \mathbf{H}_k \mathbf{P}_{k|k-1} \mathbf{H}_k^T + \mathbf{R}_k,\\
\mathbf{K}_k &= \mathbf{P}_{k|k-1} \mathbf{H}_k^T \mathbf{S}_k^{-1},\\
\mathbf{\hat{x}}_{k|k} &= \mathbf{\hat{x}}_{k|k-1} + \mathbf{K}_k \mathbf{\Tilde{s}}_k,\\
\mathbf{P}_{k|k} &= (\mathbf{I} - \mathbf{K}_k \mathbf{H}_k) \mathbf{P}_{k|k-1},
\end{split}
\end{equation}
where $\mathbf{I}$ is the indentity matrix and $\mathbf{R}_{k}$ is the covariance of observation noise. From~\eqref{eq:ekf_update}, we reconstruct the quaternion and body-frame angular velocity from position, heading and rotor speed sensor measurements. 

\subsection{Anomaly Detection Details}
\subsubsection{Anomaly Detection as Hypothesis Testing}
\label{appendix:ad_ht}

Regardless of the %anomaly detector 
AD type, % of anomaly detector, 
it aims to solve a hypothesis testing problem. Given a time series sequence from sensor observation data $Y:\{Y_{k-l_0+1}, Y_{k-l_0}, ..., Y_{k-1}, Y_k\}$ of length $l_0$, the two hypotheses are:
\begin{equation}
\begin{split}
    H_0&: \text{No attack, } Y_k \text{ was received}; \\
    H_1&: \text{Under attack, } \tilde{Y}_k \text{ was received}.
\end{split}
\end{equation}
Solving the hypothesis problem is to determine whether current observation data belong to the null hypothesis distribution $H_0$, or an alternative (but {\textbf{unknown}}) distribution $H_1$. The AD maps the current observation data into a binary classification: $D(Y_k)\to \{0, 1\}$, and the detection result is: % characterized~as:
\begin{equation}
\begin{split}
    \text{True detection} &:  D(Y_k) = 1, Y_k \in H_1, \\
    \text{False alarm} &:  D(Y_k) = 1, Y_k \in H_0.
\end{split}
\end{equation}
By denoting the probability of true detection as $P_{TD}$ and %the probability of 
false alarm as $P_{FA}$, we evaluate the effectiveness of~ADs. % anomaly detector. 

To solve the hypothesis testing problem (i.e., design an AD), we design a proper mapping for $D(Y_k)\to \{0, 1\}$.

\subsubsection{Anomaly Detection for Modulated Attacks}
\label{appendix:anomaly_detection}

\begin{figure}[!t]
\centering
\includegraphics[width=0.92\columnwidth]{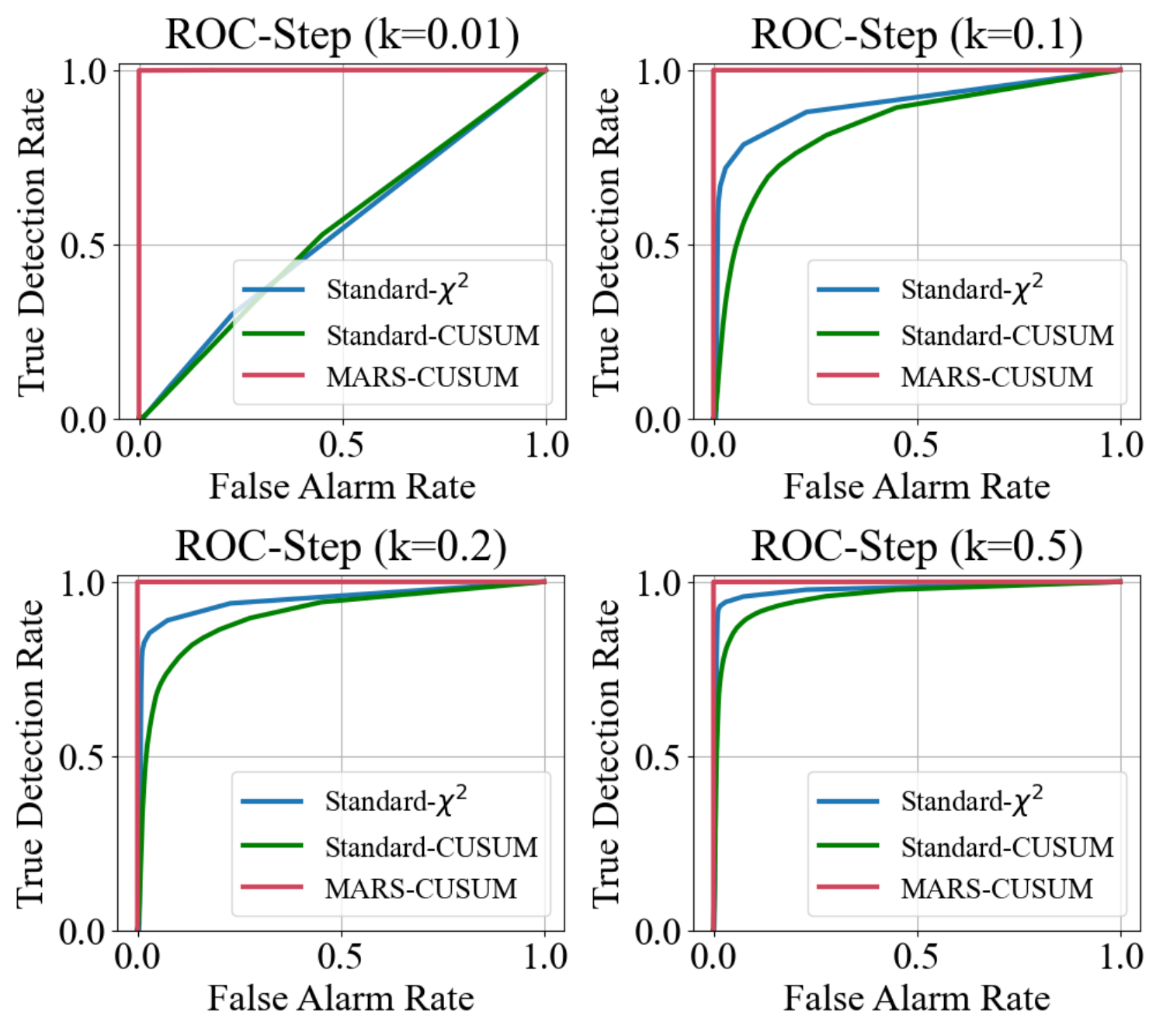}
\caption{AD offline ROC curves with Step %amplitude attack vectors.}
attacks.}
\label{fig:anomaly_detection_roc_step}
\end{figure}

\begin{figure}[!t]
\centering
\includegraphics[width=0.92\columnwidth]{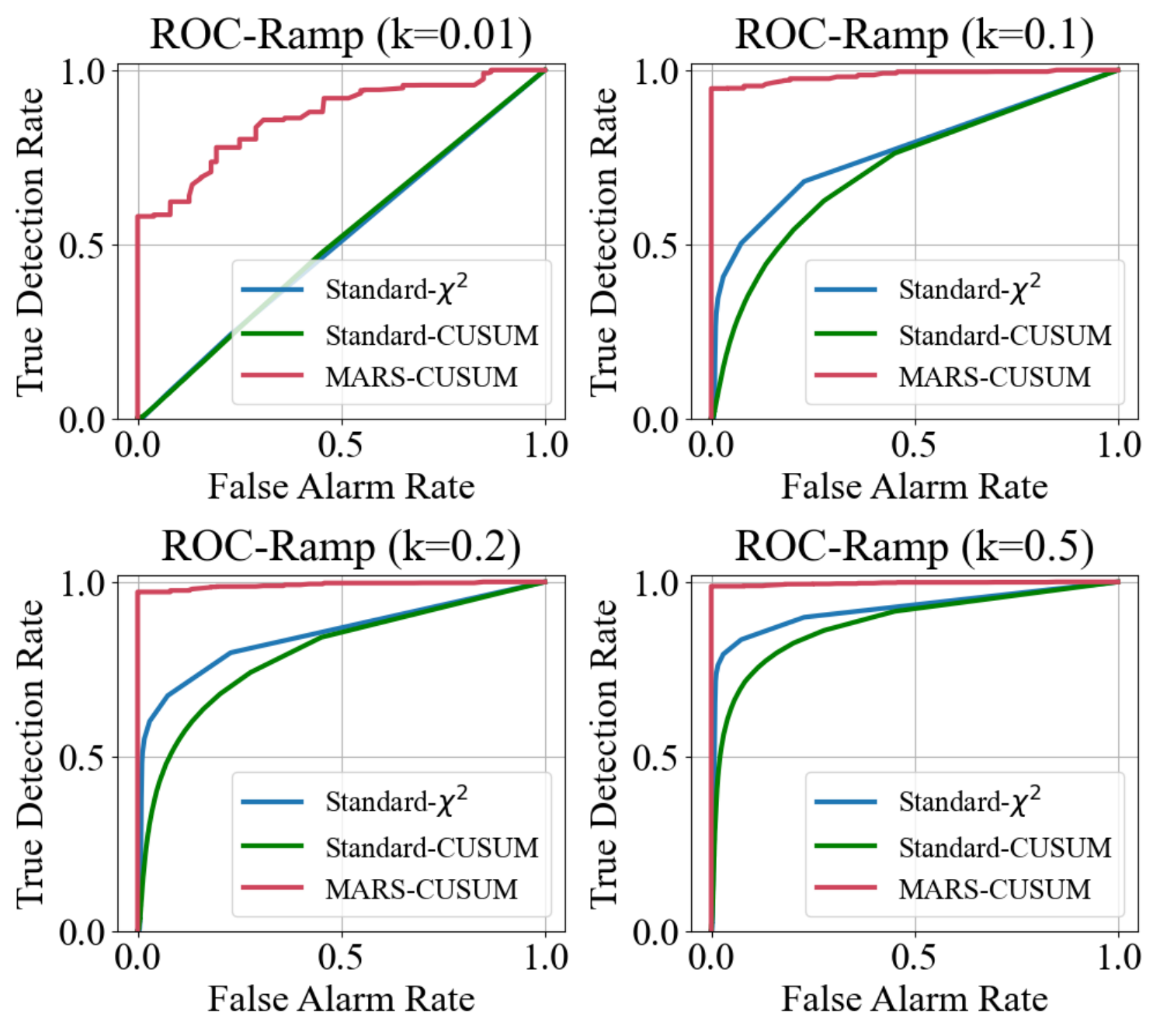}
\caption{AD offline ROC curves with Ramp %amplitude attack vectors.}
attacks.}
\label{fig:anomaly_detection_roc_ramp}
\end{figure}

% In this paper, the attacks investigated on inertial sensors are powerful and easy to detect by ADs. However, aside from 
% In addition to the previously reported physical attacks, UAVs are also susceptible to cyber attacks capable of injecting false data into the communication between sensors and control units. Therefore, we 
We also evaluate the  MARS-based AD, and compare it to the other benchmark detectors, against modulated attacks. 
Specifically, two more types of attack vectors modulated from the standard AR-DoS attacks are considered: \emph{Step Amplitude} attack and \emph{Ramp Amplitude} attack. The former can freely adjust the amplitude of the acoustic resonance while the latter can slowly increase the attack power over time. These more stealthy attacks compared to standard acoustic resonant attacks and EMI attacks pose a greater challenge to the ADs.
By choosing different scaling factors $k$, which control the maximum amplitude of the attacks, Fig.~\ref{fig:anomaly_detection_roc_step} and~\ref{fig:anomaly_detection_roc_ramp} show that the standard estimators have significant performance drop as the attack amplitude decreases. In comparison, %while MARS also may struggle as attacks become stealthy, it 
MARS significantly outperforms the standard ADs and provide more resiliency to the system. 

% \subsubsection{Step Amplitude Attacks}

% \subsubsection{Ramp Amplitude Attacks}

\subsection{MARS-PX4 Autopilot}
\subsubsection{System Parameters}
\label{appendix:system_params}

% This section elaborates on 
We now summarize the MARS-PX4 system parameters used for simulation and physical world evaluations. Table~\ref{tab:px4_sitl_sensors} and 
\ref{tab:px4_sitl_controllers} show the virtual sensors, estimators, and controller specifications used in the PX4 SITL environments, while Tables~\ref{tab:x500_sensors} and \ref{tab:x500_controllers} show the sensor configurations and firmware specifications, respectively. 
The software and firmware architecture is illustrated in Fig.~\ref{fig:mars_px4}, except for the virtual sensing setup in SITL replaced with real-world sensors. The specifications of these autopilot modules are shown in Table~\ref{tab:x500_controllers}. Specifically, the update frequency of attitude rate controller is dependent on the choice of estimators. If the IMU is not under attack, it updates at $800~Hz$ for best control performance; if MARS resilient control is activated, it adopts a more conservative $200~Hz$ update frequency in tune with the rate of rotor speed measurements to provide more stable and accurate control~signals.

\begin{table}[t!]
    % \captionsetup{skip=10pt}
    \caption{PX4 SITL virtual sensors %with update frequency and data output 
    specifications.}
    \centering
    \renewcommand{\arraystretch}{1.3}
    \begin{adjustbox}{width=\columnwidth}
    \begin{tabular}{c c c}
    % \hhline{===}
    \hhline{---}
    \makecell[c]{Simulation Virtual Sensors} & \makecell[c]{Update Frequency ($Hz$)} & Data Output \\ \hhline{---}
    Accelerometer & 250 & $a_x, a_y, a_z$ \\ \hline
    Gyroscope & 250 & $\Omega_x, \Omega_y, \Omega_z$\\ \hline
    Magnetometer & 250 & $M_x, M_y, M_z$\\ \hline
    Tachometer & 250 & $\omega_1, \omega_2, \omega_3, \omega_4 $\\ \hline
    GPS & 20 & $x, y, z, v_x, v_y, v_z$ \\ \hline
    Compass & 20 & $\psi$ \\ 
    % \hhline{===}
    \hhline{---}
    \end{tabular}
    \end{adjustbox}
    \label{tab:px4_sitl_sensors}
\end{table}

\begin{table}[t!]
    % \captionsetup{skip=10pt}
    \caption{PX4 SITL state estimators and control %lers with update frequency and data output specifications.
    specs.}
    \centering
    \renewcommand{\arraystretch}{1.3}
    \begin{adjustbox}{width=\columnwidth}
    \begin{tabular}{c c c}
    % \hhline{===}
    \hhline{---}
    \makecell[c]{Estimators \& Controllers} & \makecell[c]{Update Frequency ($Hz$)} & Data Output \\ \hhline{---}
    PX4-Standard State Estimator & 250 & $\hat{q}, \hat{x}, \hat{y}, \hat{z}, \hat{v}_x, \hat{v}_y, \hat{v}_z $ \\ \hline
    MARS-Resilient State Estimator & 250 & $\hat{q}, \hat{\Omega}, \hat{x}, \hat{y}, \hat{z}, \hat{v}_x, \hat{v}_y, \hat{v}_z$  \\  \hline
    MARS-Anomaly Detector & 250  & $\alpha, \alpha_s$ \\ \hline
    PX4-Position Controller & 50 & $q^{\text{ref}}, T^{\text{ref}}$ \\ \hline
    PX4-Attitude Controller & 250 & $\Omega_{x}^{\text{ref}}, \Omega_{y}^{\text{ref}}, \Omega_{z}^{\text{ref}}$ \\ \hline
    PX4-Attitude Rate Controller & 250 & $\omega_{1}^{\text{ref}}, \omega_{2}^{\text{\text{ref}}}, \omega_{3}^{\text{ref}}, \omega_{4}^{\text{ref}}$ \\ 
    % \hhline{===}
    \hhline{---}
    \end{tabular}
    \end{adjustbox}
    \label{tab:px4_sitl_controllers}
\end{table}

\begin{table}[t!]
    % \captionsetup{skip=10pt}
    \caption{MARS-PX4 X500 sensing specifications.}
    \centering
    \renewcommand{\arraystretch}{1.3}
    \begin{adjustbox}{width=\columnwidth}
    \begin{tabular}{c c c}
    % \hhline{===}
    \hhline{---}
    \makecell[c]{Sensors} & \makecell[c]{Update Frequency ($Hz$)} & \makecell[c]{Data Output}  \\ \hhline{---}
    IMU-Accelerometer (ICM-45686) & 800 & $a_x, a_y, a_z$ \\ \hline
    IMU-Gyroscope (ICM-45686) & 800 & $\Omega_x, \Omega_y, \Omega_z$\\ \hline
    IMU-Magnetometer (BMM150) & 800 & $M_x, M_y, M_z$\\ \hline
    Barometer (ICP20100) & 400 & $H_{baro}$\\ \hline
    Tachometer (TFRPM01) & 200 & $\omega_1, \omega_2, \omega_3, \omega_4 $\\ \hline
    GPS\&Compass (VICON MOCAP) & 10 & $x, y, z, v_x, v_y, v_z,\psi$ \\ \hline
    % \hhline{===}
    \hhline{---}
    \end{tabular}
    \end{adjustbox}
    \label{tab:x500_sensors}
\end{table}

\begin{table}[t!]
    % \captionsetup{skip=10pt}
    \caption{MARS-PX4 software \& firmware specifications. }
    \centering
    \renewcommand{\arraystretch}{1.3}
    \begin{adjustbox}{width=\columnwidth}
    \begin{tabular}{c c c}
    % \hhline{===}
    \hhline{---}
    \makecell[c]{MARS-PX4 Modules} & \makecell[c]{Update Frequency ($Hz$)} & \makecell[c]{Data Output} \\ \hhline{---}
    PX4-Standard State Estimator & 200 & $\hat{q}, \hat{x}, \hat{y}, \hat{z}, \hat{v}_x, \hat{v}_y, \hat{v}_z $ \\ \hline
    MARS-Resilient State Estimator & 200 & $\hat{q}, \hat{\Omega}, \hat{x}, \hat{y}, \hat{z}, \hat{v}_x, \hat{v}_y, \hat{v}_z$  \\  \hline
    MARS-Anomaly Detector & 200 & $\alpha, \alpha_s$ \\ \hline
    PX4-Position Controller & 100 & $q^{\text{ref}}, T^{\text{ref}}$ \\ \hline
    PX4-Attitude Controller & 200 & $\Omega_{x}^{\text{ref}}, \Omega_{y}^{\text{ref}}, \Omega_{z}^{\text{ref}}$  \\ \hline
    PX4-Attitude Rate Controller & 200/800 & $\omega_{1}^{\text{ref}}, \omega_{2}^{\text{\text{ref}}}, \omega_{3}^{\text{ref}}, \omega_{4}^{\text{ref}}$ \\ 
    % \hhline{===}
    \hhline{---}
    \end{tabular}
    \end{adjustbox}
    \label{tab:x500_controllers}
\end{table}

\subsubsection%{PX4 SITL Inertial Attack Profiles and Implementation Details}
{Inertial Attack Parameters}
\label{appendix:sitl_attack_profile}

\begin{figure}[!t]
    \centering
    \subfloat
    {\includegraphics[width=0.5 \columnwidth]{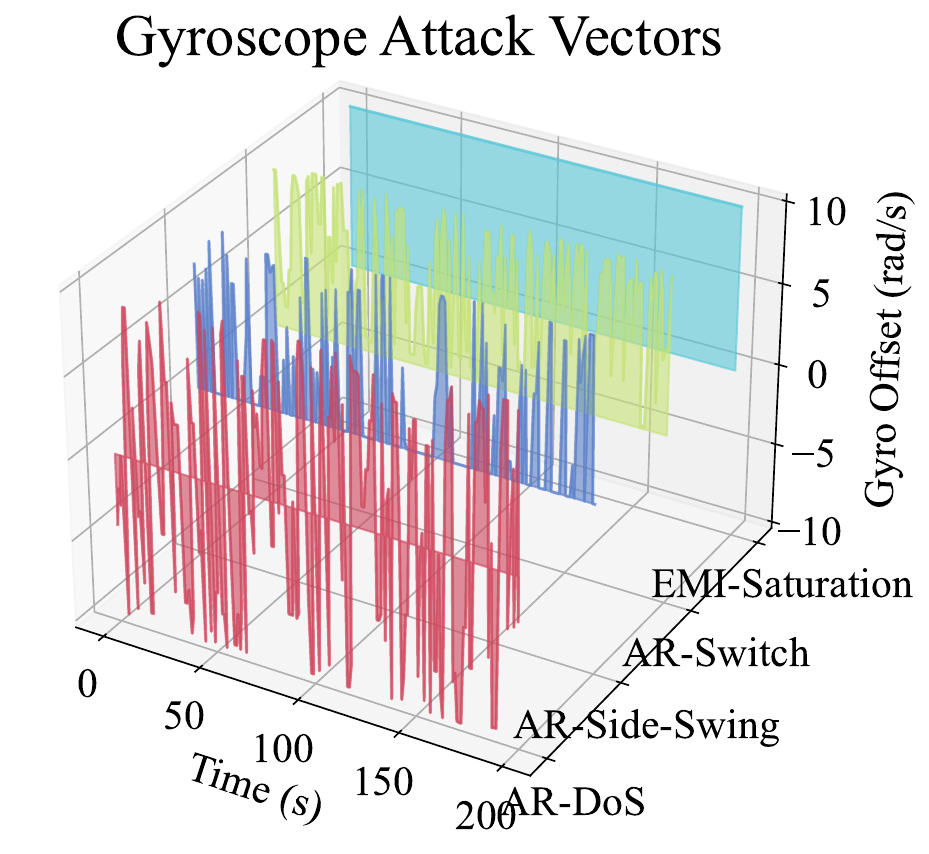}}
    \hfil
    \subfloat
    {\includegraphics[width=0.5 \columnwidth]{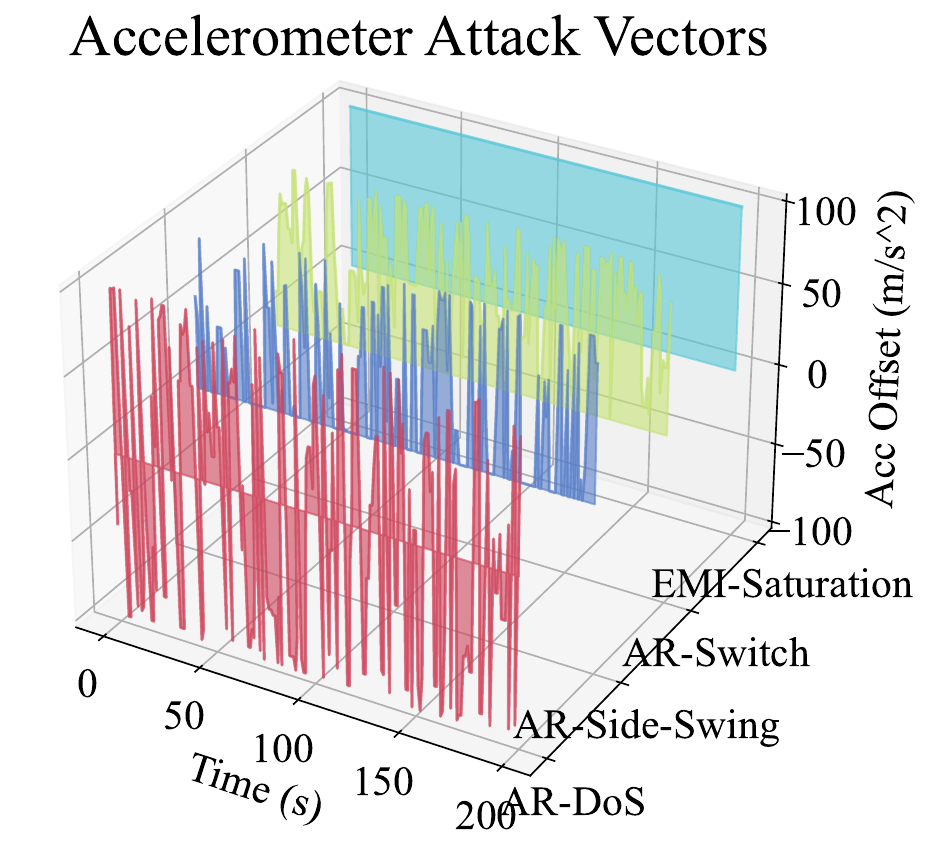}}
    \caption{Illustration of the four considered inertial sensor attack profiles for gyroscope and accelerometer.} %Left: gyroscope; right: accelerometer.}
    \label{fig:attack_profiles}
\end{figure}

In the simulated environment, we use the IMU sensor profile from the ICM-456xy MEMS Motion Sensor~\cite{ICM456XY}, with a full detection range for the accelerometer and gyroscope of $\pm~4000$ $dps$ and $\pm~32$ $g$, which correspond to $\pm~300 ~{m/s}^2$ and $\pm~70~rad/s$, respectively. We set the normal range for accelerometer and gyroscope readings at $\pm~100~{m/s}^2$ and $\pm~10~rad/s$. Thus, we set the amplitude of the acoustic resonant attack at the maximum value within the normal detection range, and the amplitude of the EMI saturation attack at the maximum value of the full range. 

The resonant frequency of the drone onboard MEMS typically ranges from $20~kHz$ to $30~kHz$, while our system sampling frequency is $250~Hz$. After the signal aliasing effect, the induced frequency lies in the range of $0$ to $125~Hz$. To select an appropriate acoustic attack induced frequency, we chose $100~Hz$ and introduced a sampling drift of $500$ $\mu$s to simulate attack signal dispersion in real~sensors. 

The attack vectors are illustrated in Fig.~\ref{fig:attack_profiles}. Also, four PX4 SITL inertial attack profiles and their corresponding configurations are summarized in Table~\ref{tab:px4_attack_vectors}.

\begin{table}[t!]
    % \captionsetup{skip=10pt}
    \caption{PX4 SITL inertial attack profiles with implementation details for acoustic resonant attack (AR) and electromagnetic interference attack (EMI).}
    \centering
    \renewcommand{\arraystretch}{1.3}
    \begin{adjustbox}{width=\columnwidth}
    \begin{tabular}{c c c c}
    % \hhline{====}
    \hhline{----}
    \makecell[c]{Inertial Sensor\\Attack Profile} & \makecell[c]{ Amplitude\\ $(A_{\text{acc}}, A_{\text{gyro}})$} & \makecell[c]{ Frequency\\ $(F_{\text{acc}}, F_{\text{gyro}})$} & \makecell[c]{Sampling Drift\\ $\sigma_{F_s}$}\\ \hhline{----}
    \makecell[c]{AR-DoS} & (100 ${m/s}^2$, 10 ${rad/s}$) & (100 $Hz$, 100 $Hz$) & 500 $\mu$s \\ \hline
    \makecell[c]{AR-Side-Swing}  & (100 ${m/s}^2$, 10 ${rad/s}$) & (100 $Hz$, 100 $Hz$)  & 500 $\mu$s \\ \hline
    \makecell[c]{AR-Switch} & (100 ${m/s}^2$, 10 ${rad/s}$) & (100 $Hz$, 100 $Hz$) & 500 $\mu$s \\ \hline
    \makecell[c]{EMI-Saturation} & (300 ${m/s}^2$, 70 ${rad/s}$) & (Inf, Inf) & Not Applicable \\ \hline
    % \hhline{====}
    \hhline{----}
    \end{tabular}
    \end{adjustbox}
    \label{tab:px4_attack_vectors}
\end{table}

\subsubsection{MARS-PX4 Computation Overhead}
\label{app:overhead}

% Quadrotor onboard flight controller is a resource-constrained computing platform, which sets strict limitations on the computational efficiency of the autopilot algorithms. In view of this tight constraint, we 

\begin{figure}[t!]
    \centering
    \subfloat{\includegraphics[width=0.503\columnwidth]{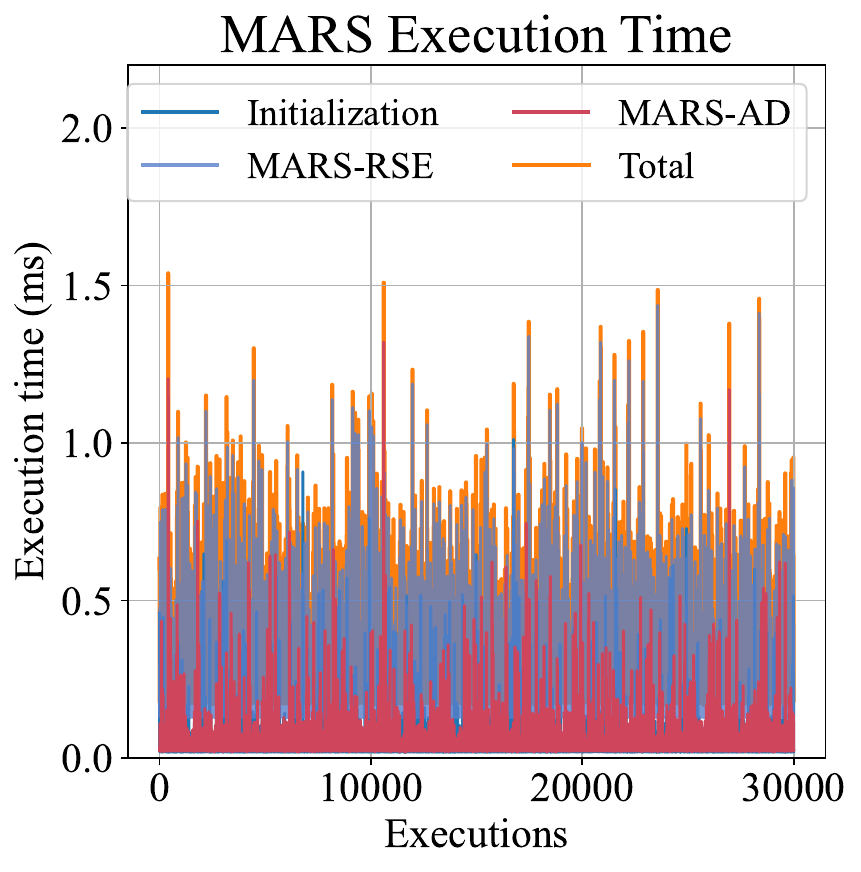}}
    \hfil
    \subfloat{\includegraphics[width=0.497\columnwidth]{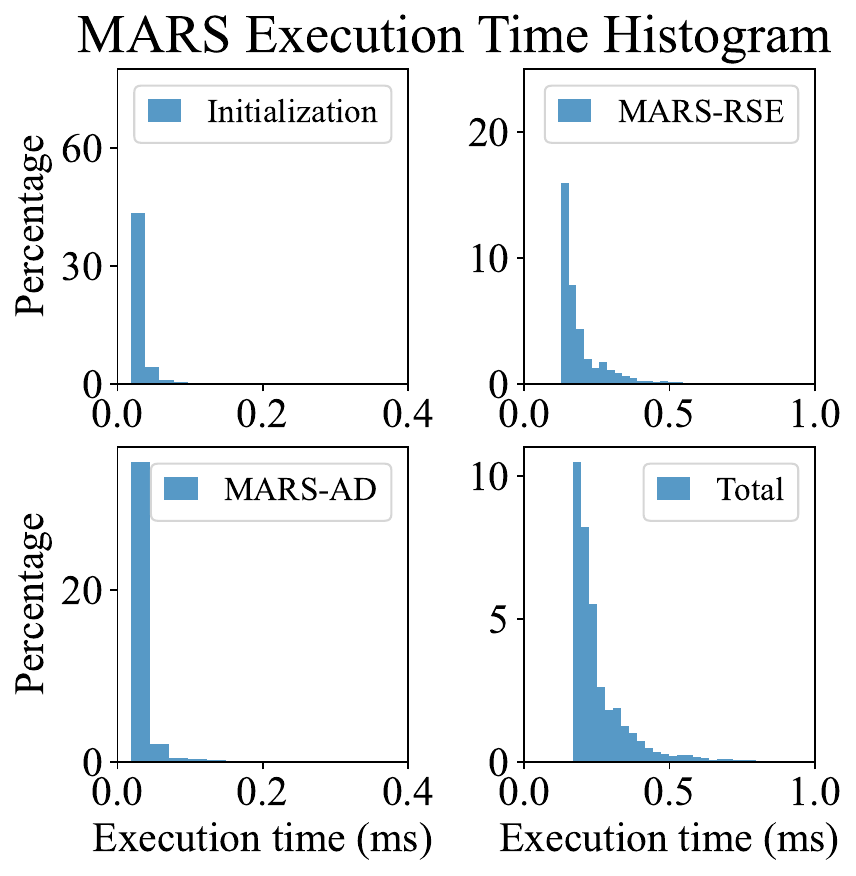}}
    \caption{MARS execution times in 10000 iterations. % at $200~Hz$. 
    Left: execution time plot; right: execution time distribution.}
    \label{fig:mars_execution_time}
\end{figure}

We measured the computation overhead of MARS modules inside the PX4 ecosystem. 
The memory usage, CPU load and module execution time are demonstrated in Table~\ref{tab:mars_cpu_and_memory_usage}.

\begin{table}[t!]
    % \captionsetup{skip=10pt}
    \caption{Computation overhead of MARS-PX4.}
    \centering
    \renewcommand{\arraystretch}{1.3}
    \begin{adjustbox}{width=\columnwidth}
    \begin{tabular}{c c c c}
    % \hhline{====}
    \hhline{----}
    \multicolumn{4}{c}{\textbf{MARS-PX4 CPU Work Queue (WQ) Usage (\%)}} \\
    \hline
     CPU Usage & WQ without MARS & WQ with MARS & Increment \\ \hline
     $\sim$47 & 0.188 & 1.499 & 1.311 \\ \hline
    \hhline{----}
    % \hhline{====}
    \multicolumn{4}{c}{\textbf{MARS-PX4 Memory Usage ($KB$)}} \\
    \hline
     Total Memory & Usage without MARS & Usage with MARS & Increment \\ \hline
     916.416 & 365.840 & 377.808 & 11.968 \\ \hline
    \hhline{----}
    % \hhline{====}
    \multicolumn{4}{c}{\textbf{MARS-PX4 Average Execution Time ($ms$)}} \\
    \hline
     Total & Initialization & MARS-RSE & MARS-AD \\ \hline
     0.261 & 0.029 & 0.200 & 0.032 \\ \hline
    % \hhline{====}
    \hhline{----}
    \end{tabular}
    \end{adjustbox}
    \label{tab:mars_cpu_and_memory_usage}
\end{table}

PX4 has a list of work queues running on CPU for efficient multi-threading. %, and we focus on the CPU usage of the specific work queue where MARS is deployed. 
We found out activating MARS does not increase the total CPU usage across all work queues (around 47\%). Within the work queue, MARS increases the CPU usage by only around 1\%. The increment in memory usage is only around $12~KB$. %, which indicate MARS only adds a tiny load on the micro-controller chip. 
At the same time, MARS is capable of finishing online processing within $0.3~ms$, which is sufficient for drone's position and attitude controllers. 

Fig.~\ref{fig:mars_execution_time} further shows the fluctuation in MARS execution time in 10000 iterations at $200~Hz$ module update frequency. It focuses on three outs main MARS components: initialization with incorporating sensor data, %MARS resilient state estimator 
MARS-RSE, %MARS anomaly detector (
MARS-AD, as well as the total execution time. The running time of MARS-RSE have spikes around every 10 executions because the update step of an EKF is executed when new observation data is received, which is the $10~Hz$ VICON-simulated GPS. %According to the execution time histogram, 
As can be seen, MARS has small oscillation in the runs and the worst case scenario is below $2~ms$, far less than the required sampling interval ($5~ms$). % of the module.

\subsubsection{Tachometer-based Estimation}
\label{app:tachometer_physical}

\begin{figure}[!t]
\centering
\includegraphics[width=\columnwidth]{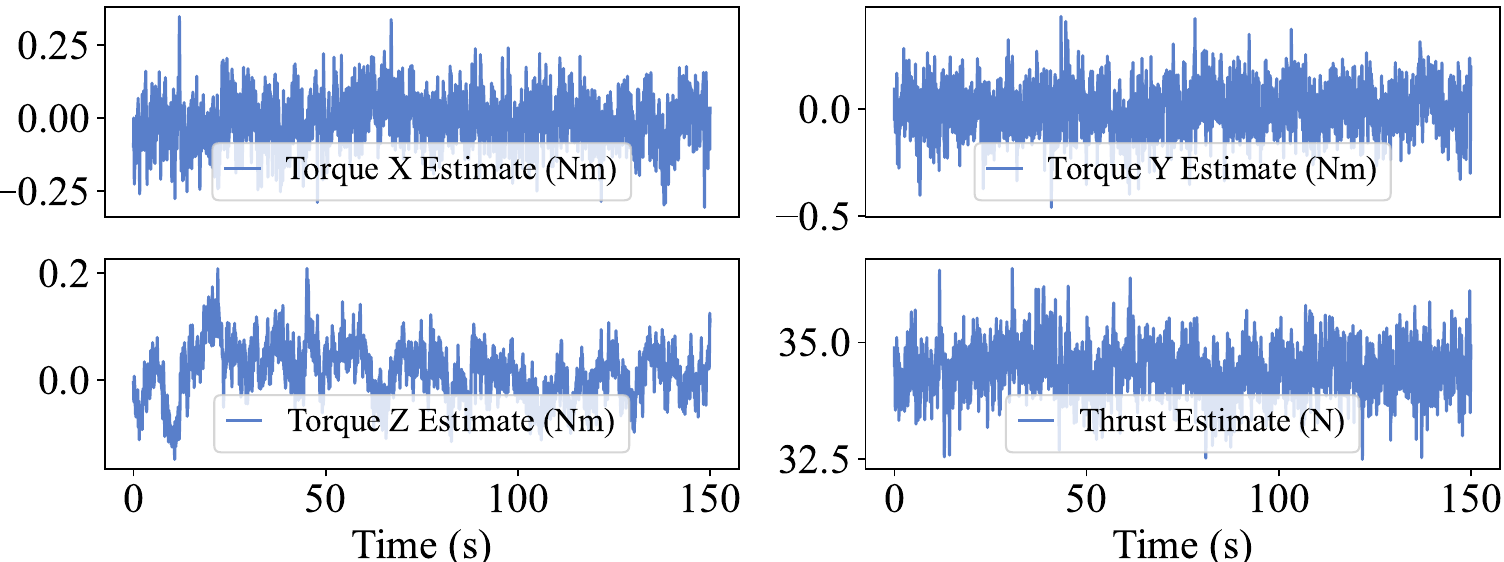}
\caption{Online torque and thrust estimates.}
\label{fig:tachometer_torque_thrust_estimate}
\end{figure}

% Tachometers measure the rate of propeller revolution by utilizing the optical flow probe that is closely mounted to the propeller base surface. 
To meet the requirement of MARS state estimator and controllers, we chose the tachometers sampling frequency at $200~Hz$ and recorded the measurements of all four rotors in real-time. 
%
% The raw tachometer sensor measurements are noisy, so we dropped the extreme values and applied a low pass filter to the raw rotor speed data %. The raw and filtered sensor measurements are shown in 
% (see Fig.~\ref{fig:tachometer_measurement}).%
%
% \begin{figure}[!t]
% \centering
% \includegraphics[width=\columnwidth]{Figures/Experiment Figures/tachometer_rotor_speed_1113_cropped.pdf}
% \caption{Tachometer rotor speed measurements before and after filtering.}
% \label{fig:tachometer_measurement}
% \end{figure}
%
As discussed in Sec.~\ref{subsec:control_input_estimate},%\todo{check now}  
we are able to estimate the UAV state %control input to the system 
by calculating the collective torque and thrust from our filtered rotor speeds and earth frame velocities. A set of proper torque compensation coefficients and bias terms for the best hovering stability is tuned experimentally as illustrated in~\eqref{eq_compensated_control_input}. The estimated torque and thrust from the filtered rotor speeds and bias compensation are shown in Fig.~\ref{fig:tachometer_torque_thrust_estimate}. The high quality of the rotor speed measurements and torque/thrust estimates lays a firm foundation for the MARS-RSE as they replace the role of inertial sensors in providing {attitude information} to the quadrotor.

\subsection{Existing IMU Attack-Recovery Methods}
\subsubsection{Learning-based Filtering Training Details}
\label{appendix:recovery_methods}

% \paragraph{Deep-Auto-Encoder (DAE)-based method~\cite{jeong2023rocking}}
\noindent\textbf{Deep-Auto-Encoder (DAE)-based method~\cite{jeong2023rocking}.}
DAE is an emerging ML-based filtering technique that excel at capturing embedded patterns in attack profiles compared to heuristic filters. %For training purposes, we 
We recorded training datasets of drone hovering and waypoint-visiting missions with randomly generated reference waypoints and headings. We then built our attack recovery dataset by adding three different variations of acoustic resonant attacks, resulting in a total of $548,000$ data samples in our training set. We did not include EMI attacks in the training set, as it is trivial for filtering-based methods to learn from saturated sensor measurements. Using the attack-recovery dataset, we trained two %deep autoencoder 
DAE models for the accelerometer and gyroscope, achieving Mean Square Error (MSE) test losses of $0.046$ and $0.005$, respectively. 
% Additionally, to ensure that our DAE model meets the $250~Hz$ update frequency required for attitude controllers, we slowed down our simulation by setting the real-time factor to $0.2$ (which is five times slower than real-time) and monitored the ROS data flow to confirm an actual $4~ ms$ model inference time.

\subsubsection{Control Benchmarks Performance in Hovering}
\label{appendix:hovering_control}

% \paragraph{Filtering-based methods struggle with un-patterned attack signals.} 
\noindent\textbf{Filtering-based methods struggle with un-patterned attack signals.} LPF could keep the drone stable for nearly $3~s$ during an AR-DoS attack but deteriorates significantly in AR-Side-Swing and AR-Switch attacks due to the faster accumulation of unfiltered in-band noise in the latter two acoustic resonant attacks. DAE does not perform better than LPF as its training involves adding noise to a clean sensor output from a normal drone flight, where the effect of the attack does not impact the drone controller. Hence, the characteristics of the attack-to-crash behaviors are not well learned. Also, DAE faces computational challenges, %as it is difficult to 
and cannot meet the low inference and communication time requirements. % before advancements in machine learning optimization on micro-controllers. 
Moreover, neither LPF nor DAE can handle sensor saturation, which results in significant control fluctuations and minimal survival time. In sensor saturation, noise is overwhelmingly dominant with little useful information left to be filtered out. Similarly, filtering-based methods struggle with other un-patterned attacks, such as incorrect values caused by unexpected data  loss, due to EMI~attacks.

% \paragraph{Attitude reconstruction is more robust in handling different attack profiles, but is limited by the quality of positioning sensors.} 
\noindent\textbf{Attitude reconstruction is more robust in handling different attack profiles, but is limited by the quality of positioning sensors.} Compared to filtering-based methods, CAF achieves much smaller control fluctuations and remains airborne longer, especially during EMI-Saturation attacks, where filtering-based methods suffer the most. However, the survival time of CAF decreases when the attack takes longer to be detected, even if the attack signal causes less impact. This phenomenon is attributed to the estimation logic of CAF, which relies on the quality of position sensor updates. It calculates roll and pitch angles geometrically at the frequency of position readings. However, it is currently challenging to access a position sensor with an update frequency comparable to that of drone attitude controllers. As a result, the effect of the attack takes longer to propagate to the position sensor, delaying CAF's ability to make necessary adjustments in response to the attack.

%% file: System_Model.tex
\subsection{Nonlinear UAV Model}
\label{appendix:drone_model}

% \subsubsection{Quadcopter Vehicle Dynamics}
% The vehicle dynamics contain critical physical constrains of the system, offering valuable insights of the logic of motion and serving as a crucial point of entry for estimating the real-time states. 

% In this paper, we focus on a quadcopter system for our analysis. The inertial frame or the earth frame is defined as the right-handed north, east, down (NED) frame, represented with $\mathcal{F}_{\mathcal{E}}$. The right-handed body frame, defined as forward, right, down (FRD) frame, is represented by $\mathcal{F}_{{\mathcal{B}}}$. The rotation from earth frame to body frame is captured in rotation matrix $\mathbf{R}(\mathbf{q})\in\mathbb{R}^{ 3\times 3}$ characterized by quaternion $\mathbf{q} =\begin{bmatrix}q_w, q_x, q_y, q_z\end{bmatrix}^T$. 

The UAV physical dynamics can be captured with a standard nonlinear quadcopter model~\cite{sun2022comparative}. The translational motion %of a quadcopter can be captured 
is described by a standard Euler-Newton equation~\cite{sun2022comparative}
\begin{equation}
\label{eq:model1}
\ddot{\mathcal{\xi}} = (\mathbf{R}\mathbf{f}^{\mathcal{B}} + \mathbf{f}_a)/m+\mathbf{g},
\end{equation}
where $\mathcal{\xi}=\begin{bmatrix}x,y,z\end{bmatrix}^T$ and $m$ denotes the center of gravity %(CoG) 
and the total mass of the vehicle, and $\mathbf{g}=\begin{bmatrix}0,0,g\end{bmatrix}^T$ is the gravity acceleration vector; $\mathbf{f}^{\mathcal{B}}=\begin{bmatrix}0,0,T\end{bmatrix}^T$ is the force vector in the body frame for the collective thrust $T$ applied at the center of mass; $\mathbf{f}_a$ is the aerodynamic drag force in the Earth frame
%
% \begin{equation}
$\mathbf{f}_a = 
\left[\begin{smallmatrix}
-k_{d,x} v_x \\
-k_{d,y} v_y \\
-k_{d,z} v_z + k_h \left(v_x^2 + v_y^2 \right)
\end{smallmatrix}\right];$
% \end{equation}
%
% \begin{equation}
% \mathbf{f}_a^B = 
% \begin{bmatrix}
% -k_{d,x} v_x^{\mathcal{B}} \\
% -k_{d,y} v_y^{\mathcal{B}} \\
% -k_{d,z} v_z^{\mathcal{B}} + k_h \left( (v_x^{\mathcal{B}})^2 + (v_y^{\mathcal{B}})^2 \right),
% \end{bmatrix}
% \end{equation}
%
here, $k_{d,x},k_{d,y}, k_{d,z},k_h$ are drag force coefficients and $v_x, v_y, v_z$ are the earth-frame velocities. 

The rotational motion can be captured as
\begin{equation}
\label{eq:model2}
\dot{\mathbf{R}}=\mathbf{R}\hat{\bm{\Omega}}^{\mathcal{B}}, \quad
\bm{\mathcal{I}}\dot{\bm{\Omega}}^{\mathcal{B}}=-\bm{\Omega}^{\mathcal{B}}\times \bm{\mathcal{I}}{\bm{\Omega}}^{\mathcal{B}} + \bm{\tau}^{\mathcal{B}},
\end{equation}
where $\bm{\Omega}^{\mathcal{B}}=\begin{bmatrix}\Omega_x,\Omega_y,\Omega_z\end{bmatrix}^T$ is the angular velocity in the body frame, $\hat{.}$ denotes the operator that maps a vector in $\mathbb{R}^3$ to a skew-symmetric matrix, $\bm{\mathcal{I}}\in \mathbb{R}^{3\times 3}$ is the inertia matrix, and
$\bm{\tau}^{\mathcal{B}}=\begin{bmatrix}\tau_x,\tau_y,\tau_z\end{bmatrix}^T$ 
is the total torque vector in the body~frame. 

% \begin{figure}[!t]
% \centerline{\includegraphics[width=\columnwidth]{Diagrams/quadcopter_frame.drawio.pdf}}
% \caption{The control input to the quadcopter system shown in FRD body frame. The four rotors generate aerodynamic forces and torques that are applied at the center of gravity of the drone.} 
% \label{fig:quadcopter_frame}
% \end{figure}

%% file: main_arxiv.bbl
\begin{thebibliography}{10}

\bibitem{martin2010true}
Philippe Martin and Erwan Sala{\"u}n.
\newblock The true role of accelerometer feedback in quadrotor control.
\newblock In {\em IEEE Int. Conf. Robot. Autom. (ICRA)}, pages 1623--1629, 2010.

\bibitem{castro2007influence}
Simon Castro, Robert Dean, Grant Roth, George~T Flowers, and Brian Grantham.
\newblock Influence of acoustic noise on the dynamic performance of mems gyroscopes.
\newblock In {\em ASME International Mechanical Engineering Congress and Exposition}, volume 43033, pages 1825--1831, 2007.

\bibitem{son2015rocking}
Yunmok Son, Hocheol Shin, Dongkwan Kim, Youngseok Park, Juhwan Noh, Kibum Choi, Jungwoo Choi, and Yongdae Kim.
\newblock Rocking drones with intentional sound noise on gyroscopic sensors.
\newblock In {\em {24th USENIX Security Symposium}}, pages 881--896, 2015.

\bibitem{kim2024systematic}
Hyungsub Kim, Rwitam Bandyopadhyay, Muslum~Ozgur Ozmen, Z~Berkay Celik, Antonio Bianchi, Yongdae Kim, and Dongyan Xu.
\newblock A systematic study of physical sensor attack hardness.
\newblock In {\em 2024 IEEE Symposium on Security and Privacy (SP)}, pages 143--143. IEEE Computer Society, 2024.

\bibitem{tu2018injected}
Yazhou Tu, Zhiqiang Lin, Insup Lee, and Xiali Hei.
\newblock Injected and delivered: Fabricating implicit control over actuation systems by spoofing inertial sensors.
\newblock In {\em 27th USENIX Security Symposium}, pages 1545--1562, 2018.

\bibitem{kune2013ghost}
Denis~Foo Kune, John Backes, Shane~S Clark, Daniel Kramer, Matthew Reynolds, Kevin Fu, Yongdae Kim, and Wenyuan Xu.
\newblock Ghost talk: Mitigating emi signal injection attacks against analog sensors.
\newblock In {\em 2013 IEEE Symp. on Security and Privacy}, pages 145--159, 2013.

\bibitem{backstrom2004susceptibility}
Mats~G Backstrom and Karl~Gunnar Lovstrand.
\newblock Susceptibility of electronic systems to high-power microwaves: Summary of test experience.
\newblock {\em IEEE Trans. on Electromagnetic Compatibility}, 46(3):396--403, 2004.

\bibitem{jang2023paralyzing}
Joon-Ha Jang, Mangi Cho, Jaehoon Kim, Dongkwan Kim, and Yongdae Kim.
\newblock Paralyzing drones via emi signal injection on sensory communication channels.
\newblock In {\em Netw. and Distributed Syst. Security Symp. (NDSS)}, 2023.

\bibitem{tu2019flight}
Zhan Tu, Fan Fei, Matthew Eagon, Dongyan Xu, and Xinyan Deng.
\newblock Flight recovery of mavs with compromised imu.
\newblock In {\em 2019 IEEE/RSJ Int. Conf. on Intelligent Robots and Systems (IROS)}, pages 3638--3644, 2019.

\bibitem{choi2020software}
Hongjun Choi, Sayali Kate, Yousra Aafer, Xiangyu Zhang, and Dongyan Xu.
\newblock Software-based realtime recovery from sensor attacks on robotic vehicles.
\newblock In {\em 23rd International Symposium on Research in Attacks, Intrusions and Defenses (RAID)}, pages 349--364, 2020.

\bibitem{fei2020learn}
Fan Fei, Zhan Tu, Dongyan Xu, and Xinyan Deng.
\newblock Learn-to-recover: Retrofitting uavs with reinforcement learning-assisted flight control under cyber-physical attacks.
\newblock In {\em IEEE ICRA}, pages 7358--7364, 2020.

\bibitem{zhang2020real}
Lin Zhang, Xin Chen, Fanxin Kong, and Alvaro~A Cardenas.
\newblock Real-time attack-recovery for cyber-physical systems using linear approximations.
\newblock In {\em 2020 IEEE Real-Time Systems Symp. (RTSS)}, pages 205--217, 2020.

\bibitem{akowuah2021recovery}
Francis Akowuah, Romesh Prasad, Carlos~Omar Espinoza, and Fanxin Kong.
\newblock Recovery-by-learning: Restoring autonomous cyber-physical systems from sensor attacks.
\newblock In {\em 2021 IEEE 27th Int. Conf. on embedded and real-time computing systems and applications (RTCSA)}, pages 61--66, 2021.

\bibitem{jeong2023rocking}
Jinseob Jeong, Dongkwan Kim, Joon-Ha Jang, Juhwan Noh, Changhun Song, and Yongdae Kim.
\newblock Un-rocking drones: Foundations of acoustic injection attacks and recovery thereof.
\newblock In {\em Netw. and Distributed Syst. Security Symp. (NDSS)}, 2023.

\bibitem{doriol2009emc}
Patrice~Joubert Doriol, Yamarita Villavicencio, Cristiano Forzan, Mario Rotigni, Giovanni Graziosi, and Davide Pandini.
\newblock Emc-aware design on a microcontroller for automotive applications.
\newblock In {\em 2009 Design, Automation \& Test in Europe}, pages 1208--1213, 2009.

\bibitem{geetha2009emi}
Shielding Geetha, KK~Satheesh~Kumar, Chepuri~RK Rao, M~Vijayan, and DC~Trivedi.
\newblock Emi shielding: Methods and materials—a review.
\newblock {\em Journal of applied polymer science}, 112(4):2073--2086, 2009.

\bibitem{karaim2019low}
Malek Karaim, Aboelmagd Noureldin, and Tashfeen~B Karamat.
\newblock Low-cost imu data denoising using savitzky-golay filters.
\newblock In {\em 2019 Int. Conf. on Comm., Signal Proc., and their Applications (ICCSPA)}, pages 1--5, 2019.

\bibitem{8836664}
Mundla Narasimhappa, Arun~D. Mahindrakar, Vitor~Campagnolo Guizilini, Marco~Henrique Terra, and Samrat~L. Sabat.
\newblock Mems-based imu drift minimization: Sage husa adaptive robust kalman filtering.
\newblock {\em IEEE Sensors Journal}, 20(1):250--260, 2020.

\bibitem{6631474}
Syed~Javed Arif, Mohammad S.~Jamil Asghar, and Adil Sarwar.
\newblock Measurement of speed and calibration of tachometers using rotating magnetic field.
\newblock {\em IEEE Trans. on Instrum. and Meas.}, 63(4):848--858, 2014.

\bibitem{ye2001anomaly}
Nong Ye and Qiang Chen.
\newblock An anomaly detection technique based on a chi-square statistic for detecting intrusions into information systems.
\newblock {\em Quality and Reliability Eng. Int.}, 17(2):105--112, 2001.

\bibitem{quinonez2020savior}
Raul Quinonez, Jairo Giraldo, Luis Salazar, Erick Bauman, Alvaro Cardenas, and Zhiqiang Lin.
\newblock Savior: Securing autonomous vehicles with robust physical invariants.
\newblock In {\em 29th USENIX Security Symp.}, pages 895--912, 2020.

\bibitem{7140074}
Lorenz Meier, Dominik Honegger, and Marc Pollefeys.
\newblock Px4: A node-based multithreaded open source robotics framework for deeply embedded platforms.
\newblock In {\em IEEE ICRA}, pages 6235--6240, 2015.

\bibitem{PX4}
{PX4 Development Team}.
\newblock Px4 official website.
\newblock \url{https://px4.io/}, 2024.

\bibitem{khan2014quadcopter}
Mohd Khan.
\newblock Quadcopter flight dynamics.
\newblock {\em International journal of scientific \& technology research}, 3(8):130--135, 2014.

\bibitem{savage1998strapdown1}
Paul~G Savage.
\newblock Strapdown inertial navigation integration algorithm design part 1: Attitude algorithms.
\newblock {\em Journal of guidance, control, and dynamics}, 21(1):19--28, 1998.

\bibitem{savage1998strapdown2}
Paul~G Savage.
\newblock Strapdown inertial navigation integration algorithm design part 2: Velocity and position algorithms.
\newblock {\em Journal of Guidance, Control, and dynamics}, 21(2):208--221, 1998.

\bibitem{xu2023sok}
Yuan Xu, Xingshuo Han, Gelei Deng, Jiwei Li, Yang Liu, and Tianwei Zhang.
\newblock Sok: Rethinking sensor spoofing attacks against robotic vehicles from a systematic view.
\newblock In {\em IEEE 8th European Symp. on Security and Privacy (EuroS\&P)}, pages 1082--1100, 2023.

\bibitem{yang2023survey}
Lei Yang, ShaoBo Li, ChuanJiang Li, AnSi Zhang, and XuDong Zhang.
\newblock A survey of unmanned aerial vehicle flight data anomaly detection: Technologies, applications, and future directions.
\newblock {\em Science China Technological Sciences}, 66(4):901--919, 2023.

\bibitem{guo2017multisensor}
Dingfei Guo, Maiying Zhong, and Donghua Zhou.
\newblock Multisensor data-fusion-based approach to airspeed measurement fault detection for unmanned aerial vehicles.
\newblock {\em IEEE Trans. Instrum. Meas.}, 67(2):317--327, 2017.

\bibitem{wang2016bias}
Jian-hong Wang, Yan-xiang Wang, and Yong-hong Zhu.
\newblock Bias compensation estimation in multi-uav formation and anomaly detection.
\newblock {\em J. Control Syst. Eng}, 4:40--50, 2016.

\bibitem{sun2017novel}
Rui Sun, Qi~Cheng, Guanyu Wang, and Washington~Yotto Ochieng.
\newblock A novel online data-driven algorithm for detecting uav navigation sensor faults.
\newblock {\em Sensors}, 17(10):2243, 2017.

\bibitem{jovanov_tac19}
I.~Jovanov and M.~Pajic.
\newblock Relaxing integrity requirements for attack-resilient cyber-physical systems.
\newblock {\em IEEE Transactions on Automatic Control}, 64(12):4843--4858, Dec 2019.

\bibitem{kwon2014analysis}
Cheolhyeon Kwon, Weiyi Liu, and Inseok Hwang.
\newblock Analysis and design of stealthy cyber attacks on unmanned aerial systems.
\newblock {\em Journal of Aerospace Information Systems}, 11(8):525--539, 2014.

\bibitem{kravchik2018detecting}
Moshe Kravchik and Asaf Shabtai.
\newblock Detecting cyber attacks in industrial control systems using convolutional neural networks.
\newblock In {\em 2018 Workshop on cyber-physical systems security and privacy}, pages 72--83, 2018.

\bibitem{Prometheus}
Prometheus - open source autonomous drone project.
\newblock \url{https://github.com/amov-lab/Prometheus}.
\newblock {A}ccessed: Nov. 13, 2024.

\bibitem{ban2013integral}
Yalong Ban, Quan Zhang, Xiaoji Niu, Wenfei Guo, Hongping Zhang, and Jingnan Liu.
\newblock How the integral operations in ins algorithms overshadow the contributions of imu signal denoising using low-pass filters.
\newblock {\em The Journal of Navigation}, 66(6):837--858, 2013.

\bibitem{MARS}
{MARS}: Defending unmanned aerial vehicles from attacks on inertial sensors with model-based anomaly detection and recovery.
\newblock \url{https://sites.google.com/view/mars-uav-recovery/home}.

\bibitem{8986669}
Wangli He, Feng Qian, Qing-Long Han, and Guanrong Chen.
\newblock Almost sure stability of nonlinear systems under random and impulsive sequential attacks.
\newblock {\em IEEE Transactions on Automatic Control}, 65(9):3879--3886, 2020.

\bibitem{DoD2012FlyingQualities}
{Department of Defense Handbook}.
\newblock {\em Flying Qualities in Piloted Aircraft}, 2012.

\bibitem{1643442}
Jeen-Shing Wang and Yen-Ping Chen.
\newblock A fully automated recurrent neural network for unknown dynamic system identification and control.
\newblock {\em IEEE Transactions on Circuits and Systems I: Regular Papers}, 53(6):1363--1372, 2006.

\bibitem{yu2017preparing}
Wenhao Yu, Jie Tan, C~Karen Liu, and Greg Turk.
\newblock Preparing for the unknown: Learning a universal policy with online system identification.
\newblock {\em arXiv preprint arXiv:1702.02453}, 2017.

\bibitem{sun2022comparative}
Sihao Sun, Angel Romero, Philipp Foehn, Elia Kaufmann, and Davide Scaramuzza.
\newblock A comparative study of nonlinear mpc and differential-flatness-based control for quadrotor agile flight.
\newblock {\em IEEE Transactions on Robotics}, 38(6):3357--3373, 2022.

\bibitem{ICM456XY}
{ICM-456XY Motion Sensor}.
\newblock \url{https://invensense.tdk.com/products/motion-tracking/6-axis/icm-456xy/}.
\newblock {A}ccessed: Nov. 13, 2024.

\end{thebibliography}
